\documentclass[sor,reprint,superscriptaddress,aps]{revtex4-1}
\usepackage{graphicx,amsmath,amssymb,color,bm}
\usepackage[normalem]{ulem}

\usepackage[bottom]{footmisc}

\usepackage{epstopdf}
\usepackage{hyperref}
\usepackage{url}
\newcommand{\be}{\begin{equation}}
\newcommand{\ee}{\end{equation}}

\newcommand{\bea}{\begin{eqnarray}}
\newcommand{\eea}{\end{eqnarray}}

\newcommand{\xg}{x_{\rm g}}
\newcommand{\kT}{k_{\rm B}T}

\newcommand{\gdot}{\ensuremath{\dot{\gamma}}}
\newcommand{\gdotmax}{\dot{\gamma}_{\rm max}} 
\newcommand{\gdotmin}{\dot{\gamma}_{\rm min}} 
\newcommand{\gdotbar}{\bar{\dot{\gamma}}} 

\newcommand{\sigmay}{\sigma_{\rm y}}

\newcommand{\Abulk}{A_\sigma}
\newcommand{\Aprofile}{A_v}

\newcommand{\nsweeps}{N}
\newcommand{\nperdecade}{n}
\newcommand{\dt}{\delta t}

\newcommand{\ogdot}{\ensuremath{\overline{\dot{\gamma}}}}
\newcommand{\osigma}{\ensuremath{\overline{\sigma}}}

\newcommand{\olambda}{\ensuremath{\overline{\lambda}}}
\newcommand{\oband}{\ensuremath{\Delta \dot{\gamma}}}
\newcommand{\aoband}{\ensuremath{\langle \Delta \dot{\gamma} \rangle}}

\newcommand{\aosigma}{\ensuremath{\langle \overline{\sigma} \rangle}}

\begin{document}

\title{Understanding rheological hysteresis in soft glassy materials}

\author{Rangarajan Radhakrishnan}%
\affiliation{Department of Physics, Durham University, South Road, Durham DH1 3LE, UK} 
\author{Thibaut Divoux}%
\affiliation{Universit\'e de Bordeaux, Centre de Recherche Paul Pascal, UPR~8641, 115 av. Dr. Schweitzer, 33600 Pessac, France}
\affiliation{MultiScale Material Science for Energy and Environment, UMI 3466, CNRS-MIT, 77 Massachusetts Avenue, Cambridge, Massachusetts 02139, USA}
\author{S\'ebastien  Manneville}%
\affiliation{Universit\'e de Lyon, ENS de Lyon, Univ Claude Bernard, CNRS, Laboratoire de Physique, F-69342 Lyon, France}
\author{Suzanne M. Fielding}%
\affiliation{Department of Physics, Durham University, South Road, Durham DH1 3LE, UK}

\date{\today}

\begin{abstract}
Motivated by recent experimental studies of rheological hysteresis in soft glassy materials, we study numerically strain rate sweeps in simple yield stress fluids and viscosity bifurcating yield stress fluids.  Our simulations of downward followed by upward strain rate sweeps, performed within fluidity models and the soft glassy rheology model, successfully capture the experimentally observed monotonic decrease of the area of the rheological hysteresis loop with sweep time in simple yield stress fluids, and the bell shaped dependence of hysteresis loop area on sweep time in viscosity bifurcating fluids.  We provide arguments explaining these two different functional forms in terms of differing tendencies of simple and viscosity bifurcating fluids to form shear-bands during the sweeps, and show that the banding behaviour captured by our simulations indeed agrees with that reported experimentally. We also discuss the difference in hysteresis behaviour between inelastic and viscoelastic fluids. Our simulations qualitatively agree with the experimental data discussed here for four different soft glassy materials. 
\end{abstract}


\maketitle

Many soft materials, including emulsions,~\cite{Becu:2006} foams,~\cite{Rouyer2008} colloids,~\cite{Mason:1995b,Knaebel:2000} microgels~\cite{Cloitre:2000} and star polymers~\cite{Rogers:2010} display exotic rheological behaviour intermediate between that of liquids and solids.~\cite{Bonnecaze:2010,Bonn:2015} At rest, or under low imposed strains, they behave as weak elastic solids and often also show ageing behaviour, in which a sample becomes progressively more solid-like as a function of the time since it was prepared.~\cite{Fielding:2000,Cloitre:2003,Negi:2010b} In contrast, for imposed stresses larger than a threshold yield stress $\sigma_\text{y}$, they show more liquid-like response, although the resulting flow might be spatially homogeneous or heterogeneous,~\cite{Ovarlez:2013b} steady or strongly time dependent,~\cite{Irani:2014} depending on the nature of the interactions between the constituent particles,~\cite{Becu:2006,Ragouilliaux:2007,Ovarlez:2009,Irani:2014} the shear history~\cite{Martin:2012,Cheddadi:2012,Kurokawa:2015} and even the boundary conditions.~\cite{Gibaud:2008,Gibaud:2009} Such phenomena have been attributed to the generic presence in these `soft glassy materials' (SGMs)~\cite{Sollich:1997,Sollich:1998} of structural disorder and metastability.  

Understanding the rheology of soft glasses remains the focus of considerable ongoing debate.~\cite{Fielding:2014,Divoux:2016} Many practical applications require the determination of the flow curve $\sigma(\dot \gamma)$, which links the shear stress $\sigma$ to the shear rate $\dot \gamma$ under conditions of a steady shear flow.  Experimentally, the measurement of this curve usually involves sweeping the shear rate (or shear stress) up or down over some prescribed interval, with some prescribed temporal duration for the sweep. While the aim is to measure the steady-state flow curve, in practice a time-dependent response is often seen. In any fluid with a fixed intrinsic relaxation timescale, the departure from steady-state can be quantified by comparing that relaxation timescale to the time of the sweep.  In a glassy material, however, the typical relaxation timescale is a complicated function of the imposed flow history. In practice, therefore, the usual strategy is to compare flow curve measurements obtained in a sequence comprising an upward followed by a downward strain rate sweep. If these two curve superimpose, the resulting master curve is taken as the steady-state flow curve. In many materials the viscous stress $\sigma-\sigma_\text{y}$ is well fit to a power-law rheology, as first noted by Herschel and Bulkley.~\cite{Herschel:1926} Often, however, hysteresis loops arise in these flow curve measurements. This phenomenon has been termed \textit{rheological hysteresis}, and has been reported widely in soft glassy materials.~\cite{Barry:1979,Chen:1992,DaCruz:2002,Holmes:2004,tenBrinke:2007,Moller:2009b,Derakhshandeh:2011,Divoux:2011b,Vasu:2013,Poumaere:2014,Kurokawa:2015,Fourmentin:2015,Mendes:2015,Perret:1996,Prestidge:1999,Labanda:2005}  

In the rheological literature, the first mention of hysteresis loops of which we are aware dates back to 1932, when E.~L.~McMillen reported a discrepancy between increasing and decreasing stress controlled ramps in paints.~\cite{McMillen:1932} H.~Green and R.~N.~Weltmann proposed computing the area enclosed inside the loop.~\cite{Green:1948}  For the materials studied in Refs.~\cite{Green:1942,Green:1948}, this loop area was found to be a decreasing function of increasing time to perform the upsweep.  K.~Nakaishi and co-workers proposed a slightly simpler protocol of sequentially increasing then decreasing the shear rate, without any pause between the two ramps.~\cite{Nakaishi:1994} For clays, they likewise found the area of the loops to be smaller for ramps performed more slowly.~\cite{Nakaishi:1997}  

Hysteresis loops are often discussed as a hallmark of rheological behaviour termed \textit{thixotropy}.~\cite{Mewis:1979}  A strict definition of the protocol used to measure the loops has so far been lacking however. For example, the sweeps are often performed over arbitrary durations compared to the material's intrinsic relaxation timescale.~\cite{Tarrega:2004,Dolz:2007,Sun:2015} Furthermore the shear rate can either first be swept from low to high values, or vice versa. Delays of arbitrary durations are often imposed at the end of the first sweep.  In 2009, J.~Mewis and N.~Wagner summed up the state of the art as follows \cite{Mewis:2009}: \textit{``The area enclosed will depend on test conditions such as the shear history prior to the start of the experiment, the maximum shear rate and the acceleration rate''.} It is also worth noting that the loop area may be computed with the flow curves plotted either in linear-linear scales or in linear-logarithmic scales. This of course strongly impacts the value of the area, because a linear abscissa gives more weight to larger shear rates.

Motivated by the aim of providing a rationalized framework for generating hysteresis loops reproducibly, and with the smallest number of free parameters, ~\citet{Divoux:2013} introduced the following experimental protocol.  First, the fluid is sheared to steady state at a high shear rate $\gdotmax$ in order to erase any previous flow history.
The shear rate is then swept from this initial high value $\gdotmax$ to a minimum value $\gdotmin$, and then back up to $\gdotmax$, without any pause between these two sweeps.  Each sweep comprises the same fixed number $\nsweeps$ of steps in the value of the shear rate, equally spaced on a logarithmic scale. The waiting time per point $\dt$ gives an equivalent sweep rate $\text{d}\log(\gdot)/\text{d}t=1/\nperdecade\dt$, where $\nperdecade=N/\log(\gdotmax/\gdotmin)$ is the number of steps per decade.  The sweep time $\dt$ is taken as the main control parameter, with $\gdotmin,\gdotmax$ and $\nsweeps$ held fixed between different experiments.  Care is taken to use large enough values of $\dt$ that a steady flow is attained at $\gdotmax$, both at the start of the downsweep and the end of the upsweep, such that any time-dependent effects are confined to the lower strain rates explored by the sweeps.

Based on the above protocol, robustly reproducible results were obtained for the area $\Abulk$ of the hysteresis loops as a function of the sweep time $\dt$. (We recall the precise definition of $\Abulk$ in Sec.~\ref{sec:protocol} below.) In particular, in Ref.~\cite{Divoux:2013}, two distinct categories of yield stress fluids were investigated: `simple' and `viscosity bifurcating' yield stress fluids. These two different types of yielding behaviours have been identified recently in the experimental literature.~\cite{Moller:2009b,Moller:2008,Fielding:2014,Divoux:2016} (In practice, however, the distinction between them may not always be sharp and there may exist a continuum of fluid types between them.) Under a sustained applied shear flow, viscosity bifurcating materials~\cite{Ragouilliaux:2007,Moller:2009b,Fall:2010b,Martin:2012} exhibit a phenomenon known as shear-banding, in which their steady-state flow field exhibits macroscopic bands of differing viscosities, with the band layer normals in the flow-gradient direction. This effect is thought to stem from a non-monotonicity in the underlying constitutive curve of shear stress as a function of shear rate (for initially homogeneous flow states). Viscosity bifurcating fluids also commonly display strong time-dependence including thixotropy or rheopexy (i.e., anti-thixotropy). In contrast, simple yield stress fluids~\cite{Coussot:2009,Ovarlez:2010,Ovarlez:2013b} show weaker time-dependence. Furthermore, their steady-state response to a sustained applied flow is one of homogeneous shear, indicating a monotonic underlying constitutive curve.

Despite this, even simple yield stress fluids often exhibit shear-banding~\cite{Divoux:2010,Divoux:2011,Divoux:2011b,Moorcroft:2011,Moorcroft:2013} during the time-dependent, transient process during which a steady flow is established out of an initial rest state, following the switch-on in a previously undeformed sample of a constant imposed shear rate $\gdot$, or a constant imposed shear stress $\sigma$.  This banding is likewise transient in nature, persisting only as long as it takes to establish a state of steady homogeneous flow, consistent with the constitutive curve of these simple yield stress fluids being monotonic. (In an oscillatory flow, however, simple yield stress fluids have recently been predicted to show sustained shear-banding as a result of the sustained time-dependence of the imposed oscillatory flow.~\cite{Ranga_2016}) 

In the same way, the rheological hysteresis observed in sweep measurements of the flow curve is often also accompanied by strongly heterogeneous, shear-banded or plug flows.~\cite{Divoux:2013,Divoux:2010} Indeed the velocity profiles themselves can show strong hysteresis, with the profile measured at any given shear rate differing between the down- and up-sweeps. In Ref.~\cite{Divoux:2013}, the area $\Aprofile$ of the hysteresis loop computed from the velocity profiles showed a robustly reproducible dependence on the sweep time $\dt$, in the same way as the area $\Abulk$ computed from the stress measurements discussed above. (We recall the precise definition of $\Aprofile$ in Sec.~\ref{sec:protocol} below.) In a carbopol microgel, which is a simple yield stress fluid, both $\Abulk$ and $\Aprofile$ were found to be monotonically decreasing functions of the sweep time $\dt$. In contrast, carbon black gels and laponite suspensions, which are strongly time-dependent yield stress fluids, display a more complex hysteresis behaviour: $\Abulk$ and $\Aprofile$ both showed a bell-shaped dependence on $\dt$ in laponite suspensions, and so does $\Abulk$ for carbon black gels.

Alongside the experimental literature just described, theoretical attempts to model rheological hysteresis have been relatively limited in comparison. Many studies~\cite{Cheng:1965,Toorman:1997,Zhu:2011,deSouzaMendes:2012,Petrellis:1973,Nakaishi:1997,Mewis:2009,deSouzaMendes:2012,Puisto:2015} have considered so called `fluidity models', or `structural kinetics theories'.  These typically pose a simple scalar constitutive equation for the time evolution of the shear stress $\sigma$, coupled to an equation describing the time evolution of an underlying property $\lambda$ of the fluid, which may characterise the state of its microstructure or degree of dynamical fluidity.  However many such studies fail to systematically compute the loop area as a function of the sweep time $\dt$. Many have considered an upsweep followed by a downsweep, with some performing several such sweep-cycles until the hysteresis loops attain invariance from cycle to cycle.  Mewis pointed out that a single $\lambda$-like parameter may not be sufficient to understand experimental observations and that a multi-parameter model may be necessary.~\cite{Mewis1975} Recently, Sainudiin et al.~\cite{Sainudiin2015} analysed the hysteresis in yield stress fluids using an Ising-like model. They found that the fraction of gelled sites of their model showed hysteresis under applied stress, and the area under the hysteresis loop showed a bell-shaped dependence on $\dt$.

Of the theoretical studies just described, the vast majority have simply assumed that the flow remains homogeneous. This precludes upfront the possibility of shear-banding, in notable contradiction to the experiments of Ref.~\cite{Divoux:2013}. An important exception can be found in the work of Ref.~\cite{Puisto:2015}, which studied a fluidity model of a simple yield stress fluid, coupled to the Navier-Stokes equation to allow any heterogeneity in the flow profile to be computed.  This successfully predicted rheological hysteresis and the time-dependent shear-banding associated with it.  However for viscoelastic materials it claimed a maximum in the area $\Abulk$ of the hysteresis loops as a function of the sweep time $\dt$, even for simple yield stress fluids, at odds with the experiments of Ref.~\cite{Divoux:2013}. Viscosity bifurcating fluids were not considered in Ref.~\cite{Puisto:2015}.  

The aim of the present work is to carry out a detailed theoretical study of rheological hysteresis, and the shear-banding associated with it, in both simple yield stress and viscosity bifurcating fluids, and to support this with more detailed experimental results than were presented in Ref.~\cite{Divoux:2013}.  Our calculations are performed within a generalised fluidity model and the soft glassy rheology model.~\cite{Fielding:2014} Both have been used previously to study transient shear-banding in other time dependent protocols.~\cite{Moorcroft:2011,Moorcroft:2013,Ranga_2016}  

An important finding of our work is that, for simple yield stress fluids, the areas $\Abulk$ and $\Aprofile$ of the hysteresis loops measured from bulk rheology and flow velocimetry are always predicted to decrease monotonically with increasing sweep time $\dt$, at least in any regime where $\dt$ is long enough to ensure a steady-state at the maximum shear rate $\gdotmax$ (which was always ensured in the experiments of Ref.~\cite{Divoux:2013}). We also offer an explanation for the claimed non-monotonic dependence of $\Abulk$ and $\Aprofile$ on $\dt$ for simple yield stress fluids in Ref.~\cite{Puisto:2015}, in contradiction to the experiments of Ref.~\cite{Divoux:2013}.  Our other main contribution is to show that a proper consideration of underlying steady-state shear-banding is crucial to understanding the bell shaped curves $\Abulk(\dt)$ and $\Aprofile(\dt)$ of hysteresis loop area as a function of sweep time seen experimentally in viscosity bifurcating yield stress fluids.~\cite{Divoux:2013}  

The paper is structured as follows. In Secs.~\ref{sec:experiment} and~\ref{sec:theory} respectively we outline our experimental and theoretical methods. In Sec.~\ref{sec:syf} we present our results for simple yield stress fluids by focusing first on experiments and second on theoretical calculations. Results are presented in the same sequence for viscosity bifurcating fluids in Sec.~\ref{sec:vb}. Because of the simplified and generic nature of the fluidity and SGR models used here, we do not attempt a detailed quantitative comparison between theory and any particular set of experimental data. Rather, we aim to capture theoretically the salient qualitative features of the experimental observations, in particular regarding the dependence of the hysteresis loop area on sweep time, separately for simple yield stress and viscosity bifurcating fluids. In Sec.~\ref{sec:conclusions} we present our conclusions and perspectives for future work.  

\section{Experimental Methods}
\label{sec:experiment}

\subsection{Rheology coupled to velocimetry}

Experiments are performed in a Couette geometry made of Plexiglas with a rotating inner cylinder of radius 24~mm and  height 28~mm, and a fixed outer cylinder of radius 25~mm, yielding a gap $e=1$~mm. The boundary conditions are either smooth (polished Plexiglas of typical roughness 15~nm) or rough (sandblasted Plexiglas of typical roughness 1~$\mu$m). The cell is topped by a homemade lid to minimize evaporation. Rheological data are recorded with a stress-controlled rheometer (MCR 301, Anton Paar). As recalled in the introduction, two flow curves are successively measured through the same protocol as in Ref.~\cite{Divoux:2013}: we first decrease the shear rate $\dot \gamma$ from high shear ($\dot \gamma_\text{max}=10^3~$s$^{-1}$) to low shear ($\dot \gamma_\text{min}=10^{-3}~$s$^{-1}$) through $N=90$ successive logarithmically spaced steps of duration $\delta t$ each, and then immediately increase $\dot \gamma$ back from $\dot \gamma_\text{min}$ up to the initial value $\dot \gamma_\text{max}$ following the same $N$ steps in reverse order. Starting from high shear rates ensures that the sample is fully fluidized and that the initial stage of the experiment is reproducible, independent of any previous shear history. In general, the downward and upward flow curves, $\sigma_\text{down}(\dot \gamma)$ and $\sigma_\text{up}(\dot \gamma)$, do not coincide and define a hysteresis loop.

Simultaneously to the flow curves, the azimuthal velocity $v$ is measured as a function of the radial distance $r$ to the rotor, at about 15~mm from the cell bottom, and with a spatial resolution of 40~$\mu$m by means of ultrasonic velocimetry.~\cite{Manneville:2004a} A piezo-polymer immersion transducer of central frequency $f=36$~MHz (Panametrics PI 50-2) emits short ultrasonic pulses and collects the resulting pressure signals backscattered by the sample seeded with a small quantity of hollow glass spheres. For a given pulse, the backscattered pressure signal, referred to as the ultrasonic speckle signal, results from the interferences of all the waves scattered by the glass spheres in the thin cylindrical acoustic beam (of diameter $\sim60~\mu$m) during the propagation of the pulse across the gap of the Couette cell. In such a speckle signal, echoes that arrive at a given time correspond to a given propagation distance from the transducer to the scatterers and back to the transducer. Therefore, this speckle signal reflects the position of the scatterers for the corresponding pulse. As scatterers are advected with the sample, the cross-correlation of two successive speckle signals over small time windows gives access to the local velocity $v(r,t)$ of the sample. The pulse repetition frequency (PRF), i.e. the time interval between two pulses, is set proportional to the shear rate, typically $f_\text{PRF}=50\dot\gamma$ for a 1-mm gap, so that the maximum displacement between two ultrasonic pulses remains smaller than half an acoustic wavelength. This rheo-ultrasonic setup allows us to access velocities from $1~\mu$m\,s$^{-1}$ to 1~m\,s$^{-1}$. The reader is referred to Ref.~\cite{Manneville:2004a} for full technical details about ultrasonic data acquisition and processing.

Here, the shear rate is swept over six orders of magnitude for each flow curve with waiting times as small as 1~s per point. We thus need to adjust the PRF to the current value of the shear rate. To do this, we use the analog output of the MCR 301 rheometer to monitor the shear rate in real time: a standard data acquisition board (NI USB 6009, National Instruments) digitizes the shear rate signal which is read by the Labview program that controls ultrasonic velocimetry and constantly updates the PRF according to the measured $\dot \gamma$.

\subsection{Experimental systems}

We study four different complex fluids: ($i$) a carbopol microgel (ETD~2050) at 1\%~wt. which is neutralized by sodium hydroxide under magnetic stirring at 300~rpm,~\cite{Divoux:2011} ($ii$) a commercial mayonnaise (Casino), i.e. a dense emulsion, ($iii$) a 2.5\%~wt. clay suspension obtained by mixing laponite powder (Rockwood, grade RD) with ultrapure water,~\cite{Gibaud:2009} and ($iv$) a carbon black gel prepared in the absence of any dispersant by mixing CB particles (Cabot Vulcan XC72R of density 1.8) in a light mineral oil (from Sigma, density 0.838, viscosity 20~mPa\,s) at weight concentration 8\%~wt. \cite{Gibaud:2010,Grenard:2014}

In order to provide ultrasonic scattering, samples ($i$), ($iii$) and ($iv$) are seeded respectively with 1\%, 0.3\% and 1\%~wt. hollow glass microspheres of mean diameter 6~$\mu$m (Sphericel, Potters) and density 1.1 g\,cm$^{-3}$. The droplet size in sample ($ii$) is well adapted to scatter ultrasound efficiently so that seeding the fluid with acoustic tracers is not required in this case.

Carbopol microgels and concentrated emulsions are generally considered as simple yield stress fluids~\cite{Moller:2009b} so that experimental results obtained on samples ($i$) and ($ii$) will be discussed together in Sect.~\ref{sec:exp_syf}. On the other hand, laponite suspensions and carbon black are known to be strongly time-dependent systems. The former was shown to display a viscosity bifurcation~\cite{Coussot:2002b,Gibaud:2009} while the latter is a rheopectic yield stress fluid that shows delayed yielding.~\cite{Gibaud:2010,Ovarlez:2013} Both samples ($iii$) and ($iv$) are discussed below in Sect.~\ref{sec:exp_vb}.

In order to clearly distinguish the response of these different fluids, we find it necessary to present some experimental results that have previously been published. Figs.~\ref{fig:carbopol} (b) and (c), ~\ref{fig:mayo}(b), ~\ref{fig:laponite} (b) and (c), ~\ref{fig:CB}(b), and ~\ref{fig:several} (a) and (c) of this work contain experimental data presented in Ref.~\cite{Divoux:2013} and its associated Supplementary Material. The other experimental data presented in this work have not previously been published.

\section{Theoretical models and  methods}
\label{sec:theory}

\subsection{Flow geometry}
\label{sec:geometry}

We study shear flow between infinite parallel plates at $y=0,L$, with the upper plate moving in the positive $x$ direction at a speed $\gdotbar(t) L$ at any time $t$, giving a shear rate $\gdotbar(t)$ spatially averaged across the sample. We return in Sec.~\ref{sec:protocol} to prescribe the time-dependent function $\gdotbar(t)$ in the protocol of interest here. The method of our calculation essentially assumes the plates to be flat.  As described in the next section, however, we also include a simplified toy description of a small curvature of the flow cell, with its associated slight stress gradient across the gap between the plates.  

\subsection{Force balance}
\label{sec:force}

We adopt a one-dimensional description in which the flow variables are allowed to vary only in the flow-gradient direction $y$ across the rheometer gap, with translational invariance assumed in the flow direction $x$ and vorticity direction $z$. We use a simplified scalar approach in which we consider only the shear component of the stress tensor, neglecting any normal stress components.

We assume the total stress $\sigma(y,t)$ on a streamline at position $y$ at time $t$ to comprise the sum of a viscoelastic contribution $\Sigma(y,t)$ from the fluid microstructure (emulsion droplets, say) and a Newtonian solvent contribution of viscosity $\eta$:
\begin{equation}
  \sigma(y,t)=\Sigma(y,t)+\eta \dot{\gamma}(y,t).
  \label{eq:tot_stress}
\end{equation}
We work throughout at zero Reynolds number, assuming conditions of creeping flow. For flow between strictly flat parallel plates, the total stress field $\sigma$ would then have to be spatially uniform, independent of position $y$. However, as noted above we also choose to mimic in a simplified way a slight stress variation across the gap between the rheometer plates, as would arise in a curved Couette, cone-plate or plate-plate device.  To do so we assign a slight spatial dependence to the total stress field:
\begin{equation}
  \sigma(y,t)=\osigma(t)\left[1+\kappa \cos(\pi y/L)\right],
  \label{eq:constraint}
\end{equation}
characterised by a small heterogeneity parameter $\kappa$. The spatial average $\sigma(y,t)$ over $y$ is denoted $\bar{\sigma}(t)$. 

Although this does not give a faithful representation of the true stress dependence in any actual curved flow device, it does capture the essential effect of cell curvature for our purposes, which is to give a slight heterogeneity to the stress field. This acts as a small perturbation that can trigger the onset of a shear-banding instability in an initially homogeneous flow, and also plays a role in determining the position of shear-bands across the gap. 

\subsection{Constitutive models}
\label{sec:constitutive}

The dynamics of the viscoelastic stress $\sigma$ in Eq.~\eqref{eq:tot_stress} is specified by a constitutive model. In order to address the experimental observations of Ref.~\cite{Divoux:2013} we perform calculations within a fluidity model and the SGR model. 

\subsubsection{Fluidity model}
\label{sec:fluidity}
Following many other fluidity models,~\cite{Mewis:2009,deSouzaMendes:2012,Armstrong:2016} we adopt a single Maxwell mode for the dynamics of the viscoelastic stress:
\begin{equation}
  \frac{d\Sigma}{dt}=G_0\left[\gdot -\frac{\Sigma}{\eta_0\lambda}\right].
	\label{eq:sig_m}
\end{equation}
The first term in the bracket on the right hand side (RHS) describes elastic loading with flow, and the second describes viscoelastic relaxation. The effective viscoelastic relaxation timescale $\tau=\eta_0\lambda/G_0=\lambda \tau_0$ at any time, and the effective viscosity $\eta_0\lambda$. The elastic modulus $G_0$ is assumed constant.  Following~\cite{Moorcroft:2011,fielding2013} we assign the parameter $\lambda$ its own dynamics according to
\begin{equation}
	\frac{d\lambda}{dt} =r\left[1  - \frac{\lambda}{1+1/f(\dot{\gamma})}\right]  +r d^2 \nabla^2\lambda.
	\label{eq:phi_m}
\end{equation}
For a simple yield stress fluid, we take the function
\begin{equation}
  f(\dot{\gamma}) =\frac{|\dot{\gamma}|\eta_0}{\sigmay}.
  \label{eq:f}
\end{equation}	

As noted above, the quantity $\lambda$ determines the viscoelastic relaxation timescale and effective viscosity of the fluid at any time.  Such a parameter is sometimes discussed in the literature as a measure of the `structuring' in the fluid, and sometimes of the inverse dynamical `fluidity'.~\cite{Mewis:2009} We adopt the latter nomenclature, because the rheological ageing of soft glassy materials may not be associated with any observable change in an underlying structural quantity.  However the same model can also be used to consider ageing-like processes such as gelation in which a sample does physically restructure over time.  Indeed the discussion that follows could have been cast in those terms with only minor rephrasing.

\begin{figure}[!t]
	\centering
  \includegraphics[width=8.0cm]{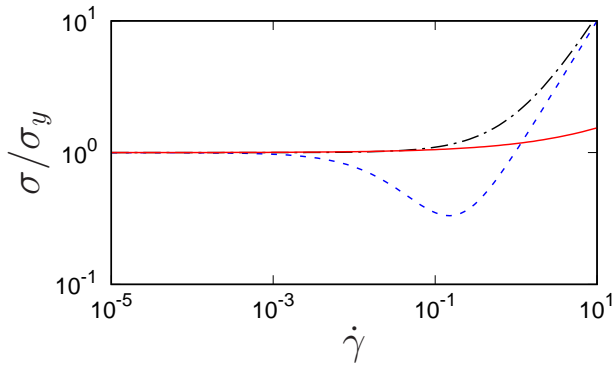}
  \caption{Underlying stationary homogeneous constitutive curves for a fluidity model of a simple yield stress fluid, $a=0$ (dot-dashed), of a viscosity bifurcating model, $p=1,a=10,\tau_1=3$ (dashed), and of the SGR model with $x=0.3$ (solid).\label{fig:constit_sketch}}
\end{figure}

In the absence of flow, the first term in the bracket on the RHS of Eq.~\eqref{eq:phi_m} captures ageing, in which the characteristic relaxation timescale of the material increases linearly as a function of the time since the sample was freshly prepared.~\cite{Fielding:2000} Within our model this ageing continues indefinitely even at long times. In contrast, some models assume a fully `structured' state at long times, with a limiting value $\lambda=1$.~\cite{Mewis:2009} The second term in the bracket accounts for rejuvenation and refluidisation by flow (often interpreted as breakdown of the underlying fluid structure).  The model could also be adapted to account for shear induced aggregation in the underlying microstructure,~\cite{Armstrong:2016} although we do not consider that possibility here.

The steady flow curve is given by $\sigma=\sigmay+(\eta_0+\eta) \gdot$, with a yield stress $\sigmay$ in the limit $\gdot\to 0$, beyond which the fluid flows with a (differential) viscosity $(\eta_0+\eta)$ as shown in Fig.~\ref{fig:constit_sketch}. The steady-state inverse fluidity is $\lambda=1+\sigmay/(\eta_0|\gdot|)$ in homogeneous flow, with the divergence in this quantity as $\gdot\to 0$ responsible for the yield stress in the flow curve. In this way, the two model parameters $\sigmay$ and $\eta_0$ determine the fluid's steady-state response.  The elastic modulus $G_0$ can instead be thought of essentially as a dynamic quantity: writing $G_0=\eta_0/\tau_0$, we recognize $G_0$ (for fixed $\eta_0$) as an effective kinetic coefficient that determines the rate of viscoelastic relaxation in Eq.~\eqref{eq:sig_m}, for any fixed value of $\lambda$. The kinetic coefficient $r$ in Eq.~\eqref{eq:phi_m} sets the rate at which the fluidity (or underlying fluid structure) adjusts to a given flow. The parameter $d$ in Eq.~\eqref{eq:phi_m} is needed to ensure a reproducible shear-banded state, with a slightly diffuse interface between the bands of characteristic width $d$.~\cite{Lu:2000,Olmsted:2000}

To model viscosity bifurcating fluids, we use Eqs.~\eqref{eq:sig_m} and~\eqref{eq:phi_m} as above, but now with the function
\begin{equation}
  f(\dot{\gamma}) =\frac{|\dot{\gamma}| \eta_0}{\sigmay} \left[1+a (\dot{\gamma} \tau_1)^{p} \right],
\label{eq:f_vb}
\end{equation}	
which for $a=0$ reduces to the function in Eq.~\eqref{eq:f} for a simple yield stress fluid.  For $a>0$ and $p>0$, the non-linearity in $f$ gives an underlying constitutive curve that has a yield stress $\sigmay$ followed by a region of negative slope, before recovering a positive slope at high shear rates. The model then admits a non-monotonic underlying constitutive curve, as shown in Fig.~\ref{fig:constit_sketch}, and allows for the steady-state shear-banding seen in viscosity bifurcating yield stress fluids.

We adopt boundary conditions at the rheometer plates of no slip for the fluid velocity together with zero-gradient for the fluidity $\lambda$. We use units in which the yield stress $\sigmay=1$, which sets our scale of stress; in which the differential viscosity of the high shear part of the flow curve is $\eta_0=1$ for simple yield stress fluids, which sets our scale of time; and in which the rheometer gap $L=1$, which sets our scale of length.  Throughout we use a small solvent viscosity $\eta=0.05$, a small characteristic width $d=0.01$ for the interface between shear-bands, and a small stress heterogeneity parameter $\kappa=0.02$. We expect our results to be robust to reasonable variations in these small quantities.

For a simple yield stress fluid this leaves as model parameters to be explored the adimensional value of $G_0$, which sets the basic scale for the rate of viscoelastic relaxation, and of $r$, which sets the basic scale for the rate of ageing (or fluidisation). All our results presented below have $r=1$, but we have checked that our main findings are robust to reasonable variations in this quantity. For a viscosity-bifurcating fluid, we use the same units and parameter values as just listed for a simple yield stress fluid, with additional parameter values $p=1$, $a=10$, and $\tau_1=3$.

The model equations are solved numerically by discretising them on a spatial grid in the $y$ direction with a mesh size $\Delta y$, and using a time-stepping algorithm with a timescale $\Delta t$. The spatially local terms are evolved using a first order implicit scheme, and the diffusive term using a backward Euler method solved by means of the Thomas algorithm.~\cite{press1996} The results presented below have $\Delta y=0.01$ and $\Delta t=5\times10^{-5}$, checked for convergence against smaller values of $\Delta y$ and $\Delta t$.  

\subsubsection{Soft glassy rheology model}
\label{sec:sgr}

The phenomenological SGR model~\cite{Sollich:1997,Sollich:1998} captures the basic glassy features of structural disorder and metastability that underpin the rheology of soft glassy materials. In its original form it addresses only homogeneous shear flows. We describe this form first, before discussing its extension to allow for heterogeneous shear-banded flows. 

In its original form the model considers an ensemble of elements that undergo activated hopping among an energy landscape of traps. Each element is taken to represent a mesoscopic cluster of a few tens of emulsion droplets or foam bubbles, say, and is assigned local continuum variables of shear strain $l$ and shear stress $kl$. These describe the cluster's state of deformation relative to a state of locally undeformed equilibrium.  Between hops, the strain of each element builds up in an affine way, following the macroscopic flow field: $\dot{l}=\gdot$. The associated elastic stress $kl$ is intermittently released by local plastic events, in which the cluster suddenly rearranges into a new local configuration, modelled as the hopping of the element out of one trap and into another.  Immediately after any hop the element selects its new trap depth randomly from a prior distribution $\rho(E)\sim\exp(-E/\xg)$, and resets its local strain $l$ to zero.

These hopping events are taken to be dynamically activated, with an element in a trap of depth $E$ and with local shear strain $l$ having a probability per unit time of yielding given by $\tau^{-1}(E,l)=\Gamma_0 \exp\left[-(E-\tfrac{1}{2}kl^2)/x\right]$.  In this way, the element's stored elastic energy $\tfrac{1}{2}kl^2$ offsets the trap depth $E$, leading to a reduced local barrier to rearrangement $E-\tfrac{1}{2}kl^2$, and exponentially fast yielding once $l=\sqrt{2E/k}$. This leads to rheological shear thinning in the fluid.  Because the typical energy barrier $E\gg \kT$, the parameter $x$ is not the true thermodynamic temperature but instead an effective noise temperature that is taken to model coupling with other yielding events elsewhere in the sample.

With the dynamics just described, the probability $P(E,l,t)$ for an element to be in a trap of depth $E$ with local shear strain $l$ evolves according to
\be
\label{eqn:master}
\dot{P}(E,l,t)+\gdot\frac{\partial P}{\partial l} = -\frac{1}{\tau(E,l)}P+\Gamma(t)\rho(E)\delta(l).
\ee
The convected derivative on the left hand side describes affine loading by shear. The first and second terms on the right hand side describe hops out of and into traps respectively, with an ensemble average hopping rate
\be
\Gamma(t)=\int dE \int dl \frac{1}{\tau(E,l)}P(E,l,t).
\ee
The macroscopic stress is defined as
\be
\sigma(t)=\int dE \int dl\; kl P(E,l,t).
\ee

The model captures a glass transition at a noise temperature $x=\xg$.  For noise temperatures $x<\xg$, the steady-state flow curve has a yield stress $\sigmay(x)$ that initially rises linearly with $\xg-x$ just below the glass point.  Beyond this yield stress the flow curve rises monotonically as $\sigma-\sigmay(x)\sim \gdot^{1-x/x_g}$. For $1<x/x_g<2$ the model displays power-law fluid behaviour with $\sigma\sim\gdot^{x/x_g-1}$. Newtonian response $\sigma\sim \gdot$ is recovered for $x/x_g>2$.  

Under low imposed strains or stresses in the glass phase, rheological ageing occurs: following sample preparation at some initial time $t=0$, the system evolves into ever deeper traps as a function of time since preparation. This confers a growing stress relaxation time $\langle \tau\rangle\sim t$, and increasingly more solid-like response as a function of the sample age.  An imposed shear of constant rate $\gdot$ can however arrest ageing and rejuvenate the sample to a steady flowing state of effective age $\langle \tau\rangle\sim 1/\gdot$.

So far we have described the model in its original form, which contains no spatial information about the location of any element and is incapable of addressing heterogeneous, shear-banded flows.~\cite{Sollich:1997,Sollich:1998} In Ref.~\cite{Fielding:2009}, we extended the model to allow for spatial variations in the flow-gradient dimension $y$.  Translational invariance is still assumed in the flow direction $x$ and vorticity direction $z$. That study also modified the model to have a non-monotonic constitutive curve, giving viscosity bifurcating YSF behaviour. In this work we remove that modification and report SGR calculations only in the case of a simple yield stress fluid. (Preliminary studies of the SGR model modified to give viscosity bifurcating behaviour however indicate very similar hysteresis phenomenology to that of the viscosity bifurcating fluidity model discussed in the previous subsection, in its viscoelastic regime. Data not shown.) 

Following~\cite{Fielding:2009}, then, we discretise the $y$ coordinate into $i=1 \cdots n_s$ streamlines of equal spacing $L/n_s$, for convenience adopting periodic boundary conditions between streamlines $i=1$ and $n_s$. Each streamline is assigned its own ensemble of $j=1\cdots m$ SGR elements, giving a shear stress $\sigma_i=(k/m)\sum_j l_{ij}$ on the $i$th streamline.

In the creeping flow conditions considered here, the shear stress must remain uniform across all streamlines in a flat flow cell. (With slight cell curvature the stress would instead be slightly heterogeneous, as mimicked by Eq.~\eqref{eq:constraint}. The extension of the procedure for imposing force balance that we shall now describe in the context of a flat cell generalises trivially to that case.) During any interval in which no hop occurs anywhere in the system, $\dot{l}=\gdot$ for every element on all streamlines and the stress change is accordingly uniform, consistent with force balance.  Supposing a hop then occurs at element $ij$ when its local strain is $l=\ell$.  (Numerically, the hopping dynamics is handled using a kinetic Monte Carlo algorithm to stochastically choose the element and time of the next hop.~\cite{Voter2007,Fielding:2009,Moorcroft:2011,Moorcroft:2013,bortz1975new,gillespie1976general}) The associated reduction in stress on that streamline would then potentially violate force balance. To correct for this, we then update all elements on the same streamline $i$ according to $l\to l + \ell/m$. This restores force balance across the streamlines, but with a stress level that has not been correctly reduced by the yielding event.  To correct for this, we further update all elements on all streamlines according to $l \to l - \ell/(m n_s)$. This algorithm can be thought of as the $\eta\to 0$ limit of imposing force balance within Eq.~\eqref{eq:tot_stress}.  For the timescales $\tau \gg \eta/k$ and flow rates $\gdot \ll k/\eta$ of interest in this work, taking this limit $\eta \to 0$ upfront gives an excellent approximation to the required physics: it amounts to assuming the effect of the Newtonian solvent to be small compared that of the viscoelastic component. 

Finally, a small stress diffusivity between neighbouring streamlines is needed to ensure a reproducible and physically realistic shear-banded state.~\cite{Lu:2000,Olmsted:2000} This is incorporated by further adjusting the strain of three randomly chosen elements on each adjacent streamline $i\pm 1$ by $\ell w(-1,+2,-1)$, after a hop on streamline $i$, where $w$ is a small dimensionless number.

We use units in which $L=1$, $\Gamma_0=1$, $k=1$ and $\xg=1$.  We take a small value $w=0.1$ for the stress diffusivity parameter. We fix the noise temperature $x/x_g=0.3$, for which the model's underlying constitutive curve is shown in Fig.~\ref{fig:constit_sketch}. Our numerical results below have $m=100$ elements on each of $n_s=100$ streamlines, checked for convergence on increasing $m$, and $n_s$.  Each run is initialized with a distribution of trap depths $P(E,t=0)=\rho(E)$, and all local stresses equal to zero, before being subject to the protocol described in the next section.

\subsection{Flow protocol and measured quantities}
\label{sec:protocol}

Within the models just described, we consider the flow curve sweep protocol introduced experimentally by Divoux et al.,~\cite{Divoux:2013} as described in the introduction above.  Before the start of the downsweep, we shear at the high initial rate $\gdotmax$ to a strain of $\overline{\gamma}=250$, which ensures a fully rejuvenated and steady initial state.  We then perform the down and upsweeps as described, taking a fixed number $\nperdecade=15$ points per decade and a fixed ratio $\gdotmax/\gdotmin=10^6$ in all our runs, giving $N=90$ sweep points in total.  Remaining parameters of the protocol are then $\gdotmax$ and $\dt$. In most of our runs we take $\gdotmax=10$ or $\gdotmax=1$, leaving $\dt$ as the parameter to be explored in detail.  However we have checked that our most important conclusions are robust to reasonable variations in this $\gdotmax$.

As discussed above, $\osigma(t)$ denotes the total stress at any time $t$ averaged across the gap between the plates, and therefore over the slight stress heterogeneity introduced by the small parameter $\kappa$ in Eq.~\eqref{eq:constraint}. At any point in the shear rate sweep we further average this quantity over the duration $\dt$ of that sweep-point, denoting the resulting average by $\aosigma$. Following Divoux et al.,~\cite{Divoux:2013} we then define the area of the stress hysteresis loop as:
\be
\label{eq:Abulk}
\Abulk=\int_{\gdotmin}^{\gdotmax}|\Delta \aosigma(\gdot)|d(\log(\gdot))
\ee
where $\Delta\aosigma$ is the difference in the stress values $\aosigma$ measured at any fixed strain rate $\gdot$ in the downsweep and the upsweep.  Although in computing this area we have chosen to use the stress $\aosigma$ time-averaged over the sweep point as just described, we could instead have taken the last measured value of $\osigma$ at any imposed $\dot{\gamma}$. Indeed we have checked that this gives qualitatively all the same behaviour as reported below using $\aosigma$, even though with slightly different numerical values. Performing the integral in Eq.~\eqref{eq:Abulk} on a logarithmic abscissa ensures equal weight to low and high shear rate regimes. Also following Divoux et al.,~\cite{Divoux:2013} we define the area in the hysteresis loop measured from the flow profiles as
\be
\label{eq:Aprofile}
\Aprofile=\int_{\gdotmin}^{\gdotmax}\int_0^L |\Delta v(\gdot,y)|dy\,d(\log(\gdot))
\ee

In reporting our results below we draw an important distinction between those runs in which the stress values of the initial point of the downsweep and the final point of the upsweep meet, such that the hysteresis loop is closed on a $\sigma,\gdot$ plot, and those in which they do not meet. This condition was ensured in all the experiments of Ref.~\cite{Divoux:2013} and we accordingly take it as a condition for any of our results to be credibly compared with experiment. However we do also report results for the hysteresis loop area even for runs in which this condition is not met. To make their distinction clear we show them as open circles in the relevant Figs.~\ref{fig:area_a0},~\ref{fig:sgr} and~\ref{fig:area_a1} .  As a working definition of the two stress values meeting, we use
\begin{equation}
  0.99 \le \langle \osigma \rangle_{\rm up}/\langle \osigma \rangle_{\rm down} \le 1.01\ .
  \label{eq:crit}
\end{equation}

To characterise the degree of shear-banding in the sample at any time we define
\begin{equation}
\label{eq:oband}
  \oband=\frac{|\dot{\gamma}_{\rm h}-\dot{\gamma}_{\rm l}|}{\ogdot}\ ,
\end{equation}
where $\dot{\gamma}_{\rm h},\dot{\gamma}_{\rm l}$ are the maximum and minimum shear rates across the cell. The time-average of this quantity over the duration of any point in the strain rate sweep is denoted $\aoband$. The fluidity of the sample spatially averaged across the rheometer gap and time-averaged over the duration of any sweep point is denoted $\langle \bar{\lambda}\rangle$.  

\section{Results: simple yield stress fluids}
\label{sec:syf}

\subsection{Experiment}
\label{sec:exp_syf}

\begin{figure}[!t]
	\centering
	\includegraphics[width=8cm]{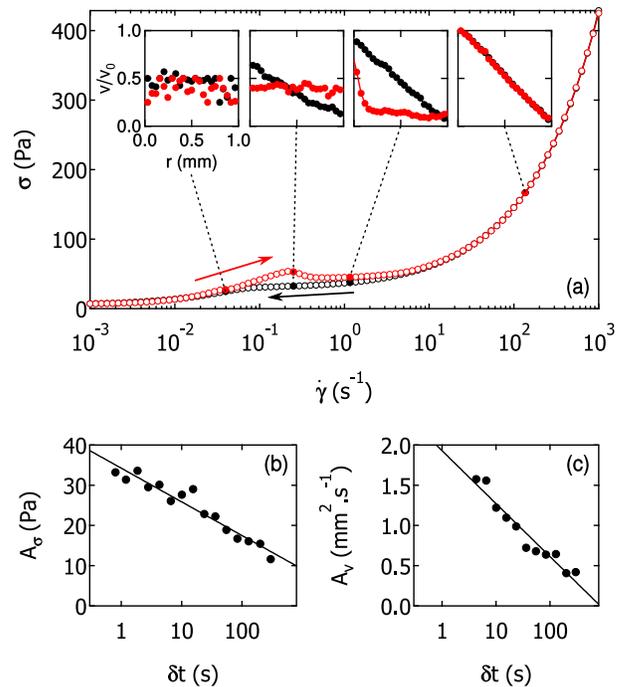}
  \caption{(a) Flow curve $\sigma$ vs. $\dot \gamma$ of a 1\%~wt. carbopol microgel obtained by first decreasing $\dot \gamma$ from 10$^3$ to 10$^{-3}$~s$^{-1}$ in 90 logarithmically spaced steps of duration $\delta t=24$~s each (black symbols), and then increasing $\dot \gamma$ over the same range (red symbols). Insets: Velocity profiles inside the gap recorded at the same shear rate during the downward (black) and upward (red) sweeps. Velocity data are normalized by the rotor velocity $v_0$ and the vertical scale goes from 0 to 1. (b) Hysteresis loop area $A_{\sigma}$ defined by Eq.~\eqref{eq:Abulk} versus $\delta t$. (c) Area $A_v$ defined from the velocity profiles by Eq.~\eqref{eq:Aprofile} versus $\delta t$. Solid lines in (b) and (c) are linear fits of the $A_{\sigma}$ and $A_v$ data in semilogarithmic scales. Experiments conducted with smooth boundary conditions (polished Plexiglas).}
\label{fig:carbopol}
\end{figure}

\begin{figure}[!h]
	\centering
  \includegraphics[width=8cm]{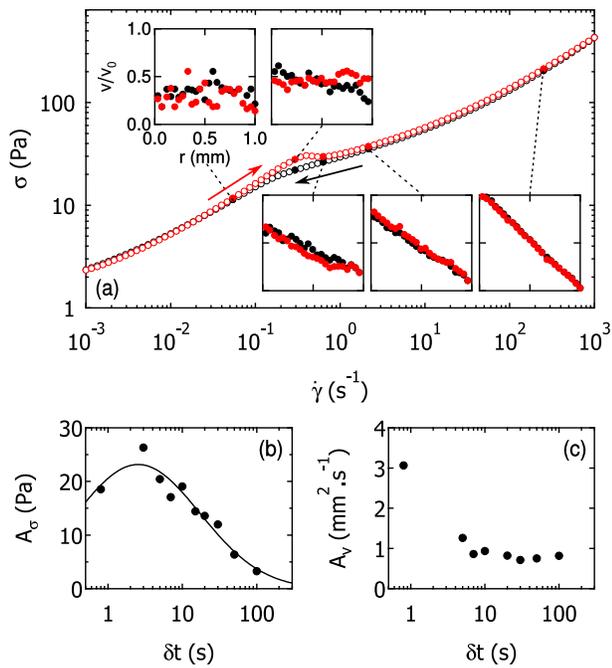}
  \caption{(a) Same as Fig.~\ref{fig:carbopol} for a commercial emulsion (Mayonnaise, Casino) with $\delta t=7$~s. The solid line in (b) serves to guide the eye through the possible maximum of $A_{\sigma}$ at the smallest $\delta t$. Experiments conducted with smooth boundary conditions (polished Plexiglas).}
\label{fig:mayo}
\end{figure}

We first examine the case of simple yield stress fluids, reporting results for a 1\%~wt. carbopol microgel [Fig.~\ref{fig:carbopol}(a)] and a commercial emulsion (mayonnaise) [Fig.~\ref{fig:mayo}(a)]. In both cases one can see that the downsweep and the upsweep almost coincide, leading to a small and yet non-negligible hysteresis loop.  In the case of the carbopol microgel, local measurements simultaneous to the rheology reveal that the velocity profiles remain homogeneous along the downward sweep, together with an ever-increasing amount of wall-slip. For shear rates smaller than $\dot \gamma \simeq 1$~s$^{-1}$ the sample is still sheared in the bulk despite large slip at the wall, while for $\dot \gamma \lesssim 0.1$~s$^{-1}$ the flow becomes plug-like (total wall-slip regime). On the upward curve, plug-like flow persists over a larger window of shear rate, namely up to $\dot \gamma \simeq 0.4$~s$^{-1}$, and gives way to heterogeneous flow profiles over one decade of shear rates, $0.4\lesssim \dot \gamma \lesssim 4$~s$^{-1}$, over which the  shear stress decreases.  This heterogeneous fluidisation of the material associated with persistent plug-like flow followed by shear-banded velocity profiles is the main contribution to the hysteresis loop. Above $\dot \gamma \simeq 4$~s$^{-1}$, the carbopol microgel flows homogeneously and both the velocity profiles and the flow curves are indistinguishable on the downsweep and on the upsweep.

The above scenario, shown here for smooth boundary conditions (polished Plexiglas), is robust to a change of $\delta t$ and does not significantly depend on the wall roughness.~\cite{Divoux:2013} Repeating the experiments for various step durations $\delta t$ allows us to compute both the rheological hysteresis loop area $A_{\sigma}$ enclosed between the upward and downward flow curves [Fig.~\ref{fig:carbopol}(b)], as well as the hysteresis loop area $A_v$ defined by the velocity profiles measured at the same shear rate during the two ramps [Fig.~\ref{fig:carbopol}(c)]. Both areas show a monotonic decreasing trend for increasing step duration. As $\delta t$ is increased, the range of shear rates over which shear-banding is observed gets narrower, leading to smaller values of $A_v$. 

In the case of the mayonnaise, shown in Fig.~\ref{fig:mayo}(a) for $\delta t=7$~s  and for smooth boundary conditions, the hysteresis in the flow curves is even smaller. Total wall-slip associated with plug-like flow is reached during the downsweep when the flow curve shows a marked change in slope at $\dot \gamma \simeq 0.3$~s$^{-1}$. The onset of flow during the upsweep coincides with a small stress overshoot around $\dot \gamma \simeq 0.5$~s$^{-1}$. This slight difference in the up and downsweep leads to a small yet measurable $A_{\sigma}$ which essentially decreases over all the range of explored $\delta t$ [Fig.~\ref{fig:mayo}(b)]. For the fastest sweeps with $\delta t=1$~s, $A_{\sigma}$ seems to increase with $\delta t$ and go through a maximum for $\delta t\simeq3$~s but it is hard to give a definite conclusion without more measurements in the low $\delta t$ range which is difficult to access experimentally.  Moreover, only homogeneous velocity profiles are detected. The non-zero hysteresis area $A_v$ results from the small mismatch, between the upsweep and the downsweep, in the shear rate corresponding to the transition from total wall-slip to a sheared state. Except maybe for the fastest sweeps, one cannot decide from Fig.~\ref{fig:mayo}(c) whether $A_v$ remains truly constant for $\delta t\gtrsim 5$~s or whether $A_v$ actually decreases below some noise level ($A_v\simeq 0.8$~mm$^2$\,s$^{-1}$) associated with the experimental uncertainty on the velocity measurements.
To conclude, the  salient feature of hysteresis loops in materials commonly classified as simple yield stress fluids is the monotonically decreasing behaviour of the loop areas $A_\sigma$ and $A_v$ down to sweeps with $\delta t\simeq1$~s. Hysteresis in the velocity measurements arises from plug-like flow and/or shear-banding following the stress overshoot in the upsweep.

\subsection{Theory}
\label{sec:th_syf}

We now study the predictions of our models of simple yield stress fluids for rheological hysteresis, considering in turn the fluidity model (with $a=0$) and the SGR model.  

\subsubsection{Fluidity model}

\begin{figure*}[!t]
	\centering
  \includegraphics[width=16cm]{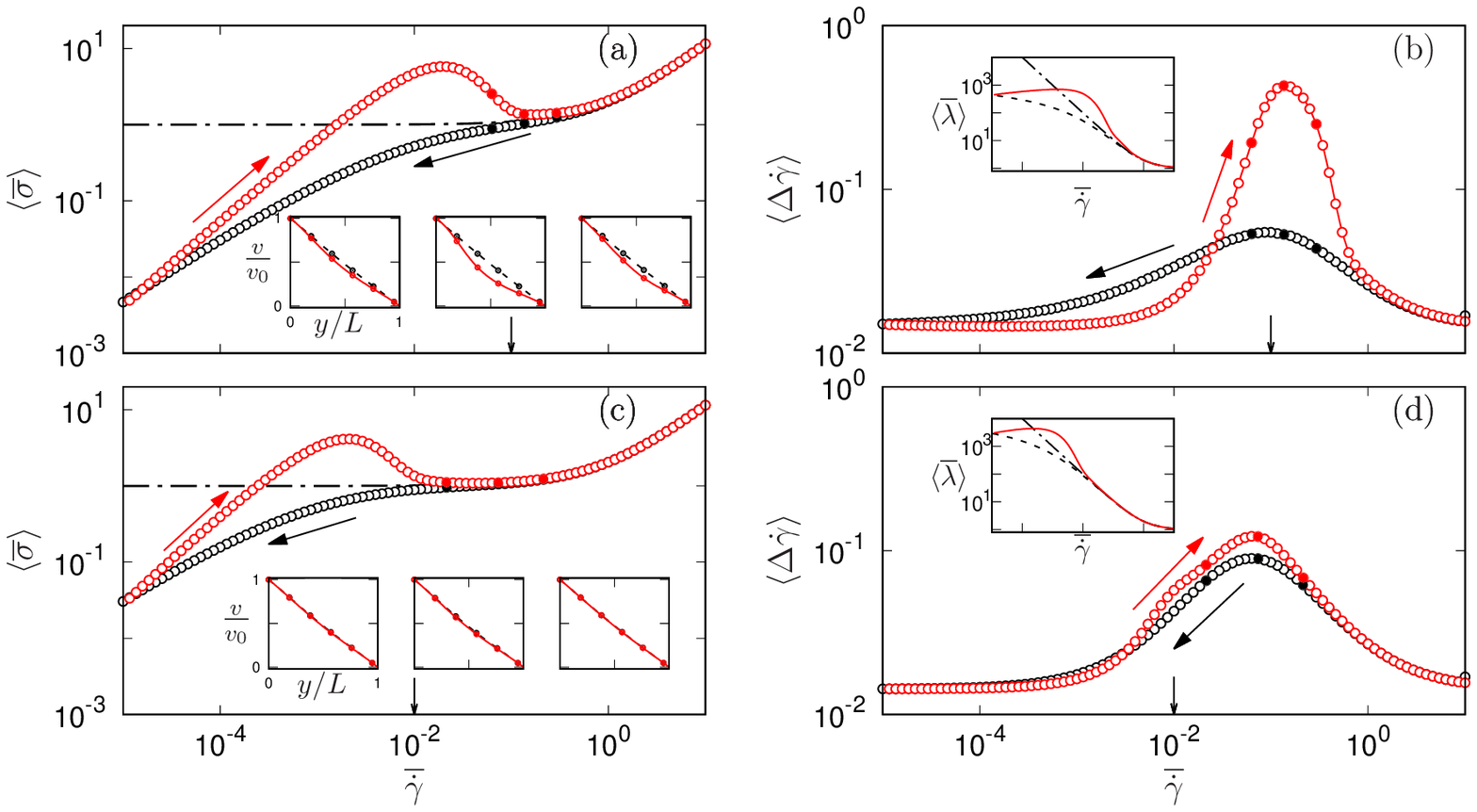}
  \caption{Simple yield stress fluidity model in the nearly inelastic regime, $G_0=100.0$, in flow curve sweeps with waiting time $\dt=10.0$ per point (top) and $\dt=100.0$ per point (bottom). In each case the left panel shows the flow curves obtained in a downsweep (black circles) followed up an upsweep (red circles), with the stationary underlying constitutive curve shown as the dot-dashed line, and the insets showing velocity profile snapshots at the times corresponding to the solid symbols in the main graph.  The right panel shows the corresponding degree of shear-banding (main graph) and the inverse fluidity (inset). The arrow pointing at the x-axis is $\ogdot=1/\dt$, which roughly coincides with the strain at which the stress first deviates from the constitutive curve during the downsweep. Videos of the rheological hysteresis in this model for $\dt=$ 10, and 100 can be found in the ESI. \footnote{\label{foot1}Electronic supplementary information (ESI) can be found here: \url{http://dx.doi.org/10.1039/C6SM02581A}.}}
  \label{fig:k_100}
\end{figure*}

The fluidity model with $a=0$ has a monotonic underlying constitutive curve, $\sigma=\sigmay+(\eta+\eta_0)\gdot$, giving simple yield stress fluid behaviour followed by a linear dependence of stress on strain rate.  Although many experimental systems show power law Herschel-Bulkley behaviour~\cite{Coussot:2014} we do not expect this to be a serious shortcoming for the purposes of this article.  

Within this model we consider the protocol proposed in Ref.~\cite{Divoux:2013}, in which the shear rate is swept from a high value $\gdotmax$ to a low value $\gdotmin=10^{-6}\gdotmax$ and back again, in each case via $N=90$ points equally spaced on a logarithmic scale, with a hold time $\dt$ per point.  This leaves to be explored the protocol parameters $\gdotmax$ and $\dt$, as well as the model parameter $G_0$ in Eq.~\eqref{eq:sig_m} and $r$ in Eq.~\eqref{eq:phi_m}. In fact all our results presented below have $\gdotmax=10.0$ and $r=1.0$, although we have checked that our main findings, in particular, concerning the overall shape of the functional dependence of the hysteresis loop area on $\dt$, are robust to varying these.  Accordingly, we focus in what follows on varying $G_0$ and $\dt$.  

Recall from Eq.~\eqref{eq:sig_m} that $G_0$ sets the degree of the material's viscoelasticity. In the limit $G_0\to\infty$ at fixed $\eta_0$, the basic timescale of viscoelastic relaxation $\eta_0/G_0\to 0$, giving an inelastic fluid.  In contrast, for $G_0=O(1)$ the material is highly viscoelastic. We shall consider one example of an essentially inelastic fluid, $G_0=100$, and one of a highly viscoelastic material, $G_0=1$, in each case exploring a wide range of values of the sweep time $\dt$.

Our results for the almost inelastic fluid, $G_0=100$, are shown in Fig.~\eqref{fig:k_100}. Although the rate of viscoelastic relaxation is fast in this case, the basic scale setting the time-frame over which the fluidity evolves is $r^{-1}=1$.  The time-evolution of the fluidity therefore interacts in a non-trivial way with the sweep time $\dt$ to give the hysteresis loops in the flow curves of panels (a) and (c).  

At the highest flow rates, the flow curves measured in the down and upsweeps coincide with each other, as well as with the stationary underlying constitutive curve $\sigma=\sigmay+(\eta+\eta_0)\gdot$, which is marked as a dot-dashed line. This has happened essentially by design: as discussed previously, we choose values of $\dt$ large enough that the flow attains a steady-state and the curves meet at the highest strain rates, to give closed hysteresis loops.  With these preliminary remarks in mind we now describe in more detail the downsweep, followed by the upsweep.  

\begin{figure*}[!t]
	\centering
  \includegraphics[width=16cm]{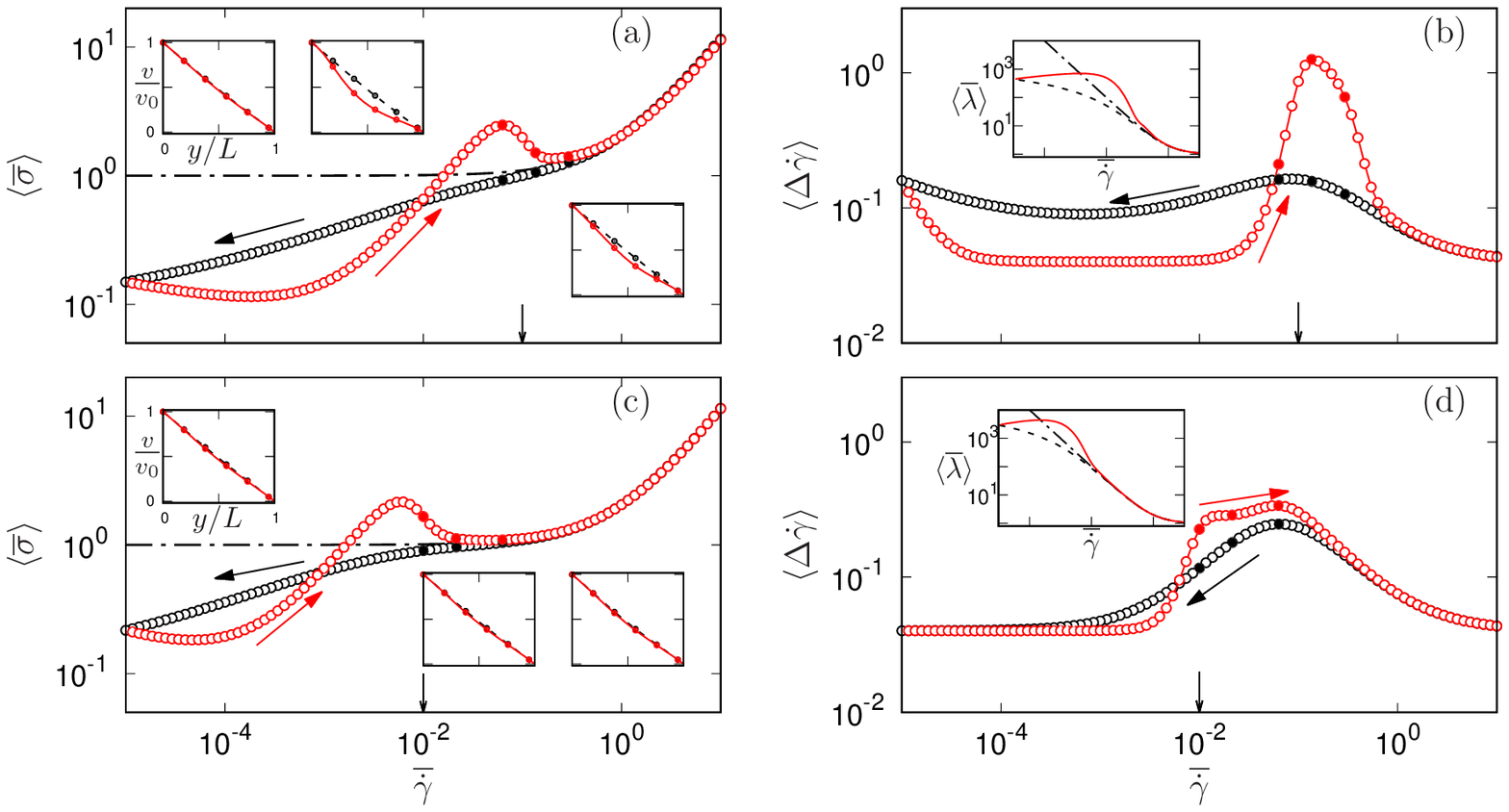}
  \caption{Simple yield stress fluidity model with high degree of viscoelasticity $G_0=1.0$, in flow curve sweeps with waiting time $\dt=10.0$ per point (top) and $\dt=100.0$ per point (bottom). In each case the left panel shows the flow curves obtained in a downsweep (black circles) followed up an upsweep (red circles), with the stationary underlying constitutive curve shown as the dot-dashed line, and the insets showing velocity profile snapshots at the times corresponding to the solid symbols in the main graph.  The right panel shows the corresponding degree of shear-banding (main graph) and the inverse fluidity (inset). The arrow pointing at the x-axis is $\ogdot=1/\dt$, which roughly coincides with the strain at which the stress first deviates from the constitutive curve during the downsweep. Videos of the rheological hysteresis in this model for $\dt=$ 10, and 100 can be found in the ESI.~\ref{foot1}} 
\label{fig:k_1}
\end{figure*}

As the strain rate is swept downwards from its initially high value, the stress departs downwards from the steady-state flow curve once the strain rate falls below a characteristic value $\ogdot \sim 1/\dt$.  The reason for this is evident from the insets to panels (b) and (d), which show the behaviour of the inverse fluidity $\lambda$ during the down and upsweeps: at these lower strain rates, $\lambda$ cannot increase quickly enough to attain its steady-state value $\olambda_{ss} = 1+1/\ogdot$ in the time $\dt$ spent at any given value of the strain rate, before the strain rate is again swept progressively lower. (The maximum rate of increase with time of $\olambda$ is linear whereas $\ogdot$ decreases exponentially.)  As a result, the sample's fluidity remains too high compared with its steady-state value.  The viscosity and shear stress accordingly remain too low compared with their values on the steady-state flow curve.  

Once the strain rate attains its minimum and is subsequently swept back upwards, the inverse fluidity and the stress initially remain too low, again failing to rise quickly enough to attain their steady-state values at any given value of the strain rate, before the strain rate is swept progressively on upwards. In contrast, a new effect is seen at higher strain rates: the inverse fluidity and stress transiently exceed their steady-state values and there is a noticeable overshoot in the stress signal at a typical strain rate that again scales as $\ogdot \sim 1/\dt$. After this overshoot the sample fluidises and the stress declines, rejoining the steady-state flow curve at the highest flow rates.

\begin{figure}[!h]
	\centering
  \includegraphics[width=8.0cm]{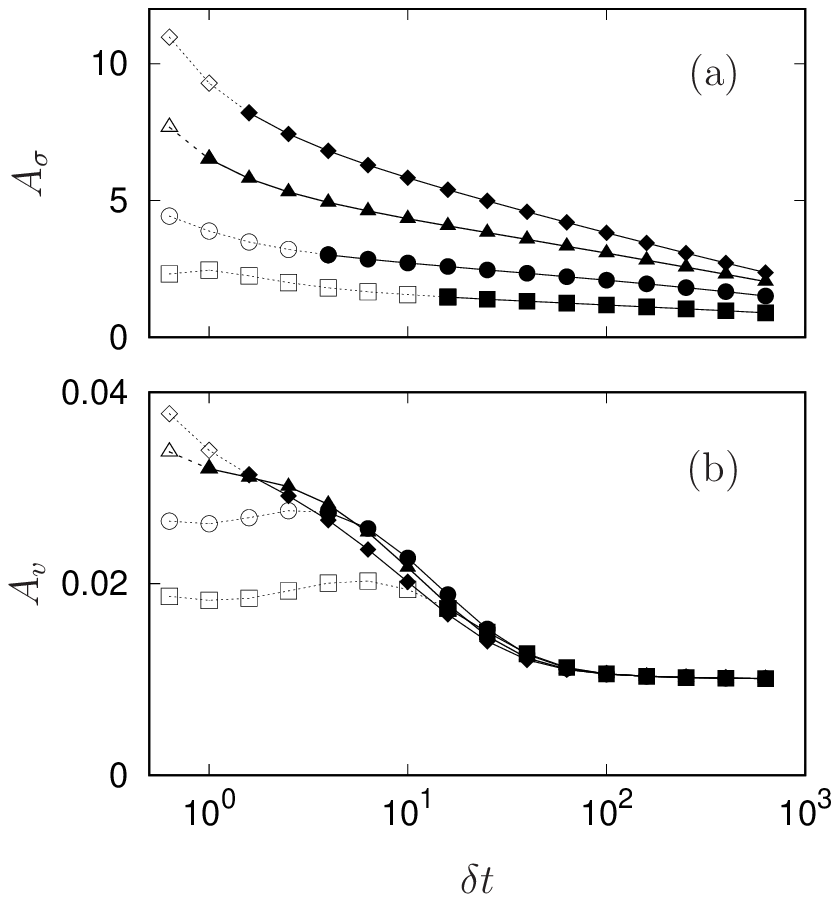}
  \caption{Area of hysteresis loops in simple yield stress fluidity model as a function of sweep time  per point $\dt$, as measured in (a) bulk flow curve measurements and (b) spatially resolved velocimetry. Curves upward in each panel correspond to decreasing levels of  viscoelasticity:  $G_0=1.0$ ($\square$), $G_0=3.16$ ($\circ$), $G_0=10.0$ ($\triangle$), and $G_0=100.0$ ($\diamondsuit$). Data points corresponding to closed hysteresis loops for which the flow curves meet at $\gdotmax$ are denoted by filled symbols; those corresponding to open loops are shown as empty symbols.
    \label{fig:area_a0}}
\end{figure}

As just described, the departure of the stress from the steady-state flow curve on the downsweep, and the overshoot in the stress before it regains the steady-state flow curve on the upsweep, both happen at characteristic flow rates $\ogdot\sim 1/\dt$.  This is seen by comparing panel (a), for which $\dt=10.0$, with its counterpart panel (c), for which $\dt=100.0$: the features just described shift to the left by a decade in moving from (a) to (c). In consequence, as the experiment is repeated for larger $\dt$ (slower sweeps), the area $\Abulk$ of the hysteresis loop decreases. This trend is shown in Fig.~\ref{fig:area_a0}(a): $\Abulk$ is a monotonically decreasing function of $\dt$ over the entire range explored.  For the larger $\dt$ values explored, the down and upsweep flow curves meet at the highest strain rates to give closed hysteresis loops, as required for a meaningful comparison with experiment.  These data points are shown as filled symbols in Fig.~\ref{fig:area_a0}(a). In contrast for the lowest $\dt$ (fastest sweeps), the flow curves fail to meet at the highest strain rates, as indicated by the open symbols.

The same model as considered here was used previously to study shear startup from rest.~\cite{Moorcroft:2011} In that work the existence of an overshoot in the stress startup signal, which typically occurs at an accumulated strain $\gamma=\mathcal{O}(1)$, was shown to lead to the transient formation of shear-bands during the subsequent process of fluidisation as the stress falls from the overshoot to its final steady-state value. Given that an upward shear rate sweep might be expected loosely to mirror a shear startup run, an obvious question is whether the stress overshoots in panels (a) and (c) are likewise associated with shear-banding.  

To explore this idea, in panels (b) and (d) we plot the degree of shear-banding $\aoband$ as a function of strain rate during the down and upsweeps, where a value $\aoband>0.5$ corresponds to an observable degree of bowing (shear-banding) in the velocity profile. For the sweep with $\dt=10.0$, a clear shear-banding effect during the upsweep in panel (b) is associated with the stress overshoot in panel (a). This is evident also in the snapshot velocity profiles in the inset of panel (a). Associated with this shear-banding effect, which is asymmetric between the up and downsweep, is a hysteresis effect in the local velocimetry measurements, as characterised by $\Aprofile$ in Eq.~\eqref{eq:Aprofile}. As can be seen in Fig.~\ref{fig:area_a0}(b), $\Aprofile$ progressively decreases with increasing $\dt$ (slower sweeps), at least in any case where the hysteresis loop is closed.  Indeed, for the sweep with $\dt=100.0$ in panels (c) and (d) of Fig.~\ref{fig:k_100}, the shear-banding effect is diminished to the point of being essentially unnoticeable. This trend is consistent with the shear startup results in Ref.~\cite{Moorcroft:2011}, where a stronger stress overshoot and more pronounced transient banding effect was associated with a faster shear startup (compared to the inverse sample age before startup commenced).

The  low background value in the shear-banding signal $\aoband$ in panels (b) and (d), below the threshold $\approx 0.5$ for a noticeable bowing in the velocity profiles, is due to the assumed presence of a slight stress heterogeneity across the rheometer gap according to Eq.~\eqref{eq:constraint}. This leads to a slight strain rate heterogeneity even in the absence of true shear-banding. The slight elevation in $\aoband$ during the downsweeps in panels (b) and (d), again of insufficient amplitude to be noticeable in the velocity profile, is essentially an artefact of our choice of normalisation in Eq.~\eqref{eq:oband}: the factor $\ogdot$ in the denominator temporarily decreases more quickly than the numerator.


We next consider a viscoelastic material, $G_0=1.0$, in which the basic timescale of stress relaxation is now comparable to the basic timescale that prescribes the ageing behaviour, such that the stress can no longer relax instantaneously for a given imposed shear rate and level of sample fluidity. As seen in Fig.~\ref{fig:k_1}, many of the overall trends in the rheological hysteresis are the same as for the inelastic fluid. However one important difference is evident: at low values of the strain rate the flow curve measured in the upsweep lies {\em below} that of the downsweep. This can be understood as follows.  In a viscoelastic material, the relaxation of stress happens gradually as a function of time. In consequence, the stress accumulated initially at the high strain rates continues to relax slowly as a function of the time during which the shear rate sweeps downwards then upward again over the low shear rate regime (where the steady-state stress value, if attainable, would be small).  Such an effect has been noted previously in works discussing a possible classification of thixotropic fluids into `ideal' and `viscoelastic'.~\cite{Mewis:2009,Larson2015}

As a result of this new effect, hysteresis now manifests itself as a double loop in each of Figs.~\ref{fig:k_1}(a) and (c), in contrast to the single loop in the counterpart panels of Figs.~\ref{fig:k_100}.  Each lobe of the double loop still contributes positively to the overall measured loop area $\Abulk$, however, due to the use of the absolute value in the definition of Eq.~\eqref{eq:Abulk}. Nonetheless, the pinching of the original single loop into a double loop has the inevitable effect of reducing $\Abulk$ for a more viscoelastic material (smaller $G_0$) compared to an inelastic fluid (larger $G_0$) in Fig.~\ref{fig:area_a0}(a).   The minimum value of $\dt$ at which the up and down sweeps meet at $\gdotmax$ to give a closed hysteresis loop (filled symbols in Fig.~\ref{fig:area_a0}) is also larger for more viscoelastic materials.  The areas $\Aprofile$ of the hysteresis loops in the velocity profile measurements vary much less with elasticity [Fig.~\ref{fig:area_a0}(b)], at least in any region where the hysteresis loops are closed (solid symbols): the shear-banding that arises during the upsweep, which is the essential ingredient in $\Aprofile$, is largely unchanged by the viscoelastic effect just described.

\subsubsection{SGR model}
\label{sec:th_sgr}
An obvious shortcoming of the fluidity model is that it only has a single Maxwell mode of stress relaxation. The SGR model provides a more realistic description of the rheology of soft glassy materials in having a power-law relaxation spectrum.

\begin{figure}[!t]
\centering
  \includegraphics[width=8.0cm]{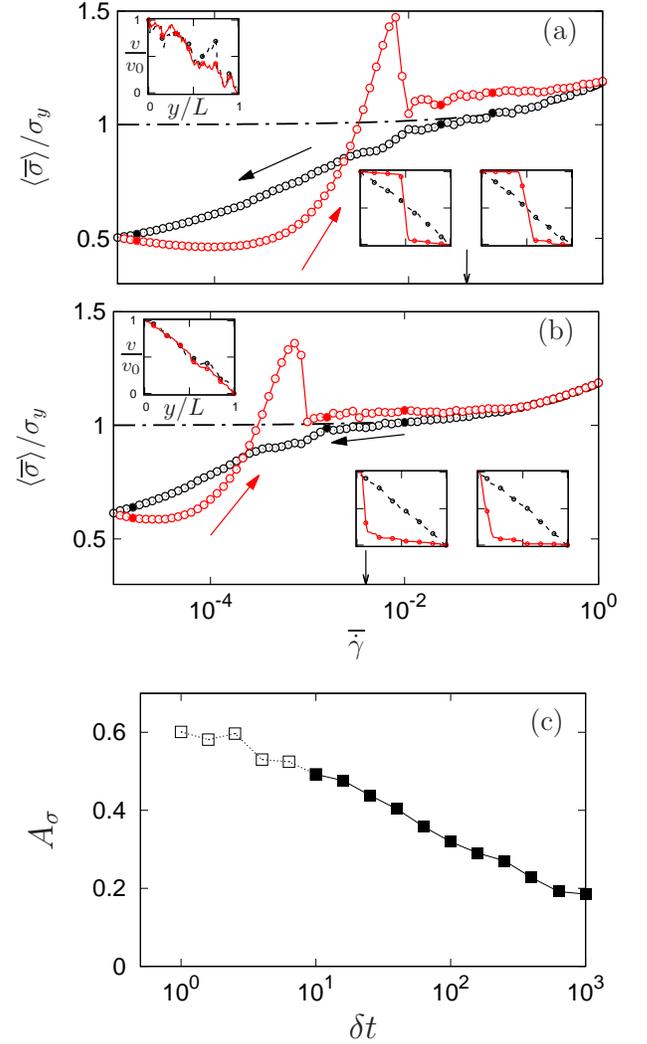}
  \caption{Flow curve sweeps in SGR model at $x=0.3$ for waiting times per point (a) $\dt=25.1$ and (b) $\dt=251$. In each case the downsweep is shown by black circles and the upsweep by red circles. Insets show snapshot velocity profiles corresponding to the data points indicated by solid symbols in the main figure. The arrow pointing at the x-axis is $\ogdot=1/\dt$, which roughly coincides with the strain at which the stress first deviates from the constitutive curve during the downsweep. The dependence of the area $\Abulk$ of the hysteresis loop area on the waiting time per point $\dt$ is shown in (c), with solid (resp.  empty) symbols corresponding to runs with closed (resp. open) hysteresis loops. Videos of the rheological hysteresis in this model for $\dt=$ 25.1, and 251 can be found in the ESI.\ref{foot1}} 
 \label{fig:sgr} 
\end{figure}

The results of our simulations of flow curve sweeps in the spatially aware form of the SGR model are shown in Fig.~\ref{fig:sgr}. As can be seen, these closely mirror the behaviour of the fluidity model.  In particular, (i) the hysteretic features in the flow curves shift to lower $\ogdot$ with increasing $\dt$, (ii) the area of the hysteresis loop $\Abulk$ decreases monotonically with increasing in $\dt$, (iii) the flow curves measured in the upsweep and downsweep do not meet at $\gdotmax$ for fast sweeps (small $\dt$) and (iv) shear-banding arises in the vicinity of the stress overshoot during the upsweep. It is especially worth noting that the SGR model mirrors most closely the case of the viscoelastic fluidity runs in Fig.~\ref{fig:k_1} in that (v) at low shear rates the flow curve measured on the upsweep lies below that of the downsweep, reflecting the slow stress relaxation of a viscoelastic material.

This close correspondence between the SGR and fluidity models gives us confidence that the phenomena reported here are generic across soft glassy materials, and not model specific.  (The quantitative details do differ between the two models, however, and we do not attempt a direct parameter match.)  It also underlines the utility of simplified fluidity models in understanding even rather complicated rheological phenomena, circumventing to some extent the need for simulating the more cumbersome SGR model.  Indeed, the computational expense of simulating the SGR model means that the velocity profiles obtained from the simulations with $m=100$ elements on each of $n_s=100$ streamlines are quite noisy. For that reason we do not show data for $\Aprofile$. We have checked, however, that the bulk signals in Fig.~\ref{fig:sgr} are robust to increasing $m$ and $n_s$.

The decrease in the area $\Abulk$ of the bulk rheological hysteresis loop with increasing sweep time $\dt$, shown in Fig.~\ref{fig:sgr}(c) for a noise temperature $x=0.3$ in the model's glass phase, also pertains for noise temperatures above the glass transition, $x>1$ (data not shown).

\subsubsection{Comparison with experiment and other theoretical work}
\label{sec:syf_comp}

Comparing the results of our simulations of the fluidity and SGR models with the experimental data for the simple yield stress fluids carbopol and mayonnaise in Figs.~\ref{fig:carbopol} and~\ref{fig:mayo}, we notice the following similarities. First, the measures $\Abulk$ and $\Aprofile$ of area in the hysteresis loops both decrease monotonically with increasing sweep time $\dt$, in both experiment and theory, excepting the leftmost data point for $\Abulk$ in mayonnaise. (We exclude from this remark the open symbols in Figs.~\ref{fig:area_a0} and~\ref{fig:sgr}(c) for the reason discussed earlier: the hysteresis loops are not closed in this case.) Second, shear-banding (or at least a bowing of the velocity profiles) arises near the stress overshoot during the upsweep, in both the experiments and theory. 

Some differences are also evident. For example, many of the experimental velocity profiles show strong wall-slip, particularly at low shear rates. Our simulations assume no slip and cannot address this.  The double hysteresis loop in the viscoelastic simulations in Fig.~\ref{fig:k_1} is either absent or weak in the experiments.  Indeed this may tell us that carbopol and mayonnaise are in fact relatively inelastic on the relevant timescales (perhaps giving behaviour intermediate between Figs.~\ref{fig:k_100} and~\ref{fig:k_1}), and independent rheological measurements to confirm this would be welcome. Alternatively, the experimental flow curves may be contaminated by wall-slip at low strain rates, although it is worth nothing that wall treatment gave only a moderate effect on the curves $\Abulk(\dt)$ and $\Aprofile(\dt)$ in Ref.~\cite{Divoux:2013}.

Recently, Puisto et al.~\cite{Puisto:2015} studied flow curve sweeps within a spatially aware fluidity model of a simple yield stress fluid. Many of our results for simple yield stress fluids are qualitatively consistent with these earlier findings. In particular, Puisto clearly demonstrated a stress overshoot during the upsweep, which is associated with shear-banding, and which shifts to lower values of strain rate with increasing sweep time $\dt$.

For inelastic fluids, Puisto et al.~\cite{Puisto:2015} predicted a monotonic decrease of the hysteresis loop areas $\Abulk$ and $\Aprofile$ with increasing sweep time $\dt$, as also reported here.  However for viscoelastic materials they demonstrated more complicated dependences, in some cases resembling the bell-shaped plots of $\Abulk(\dt)$ and $\Aprofile(\dt)$ seen experimentally in viscosity bifurcating fluids. To try to understand why, we performed a comprehensive sweep of the parameters $G_0,r,\gdotmax$ and $\dt$ in our fluidity model of a simple yield stress fluid. While indeed we found non-monotonic and bell shaped curves in some parameter regimes, in every case this non-monotonicity was confined to small values of $\dt$ (fast sweeps), for which the flow curves measured in the down and upsweeps fail to meet at high strain rates, giving open hysteresis loops that we suggest should be discarded from any comparison with experiment.  Therefore, our understanding is that in simple yield stress fluids the hysteresis loops areas $\Abulk$ and $\Aprofile$ must always decrease monotonically with increasing $\dt$ in any regime that is meaningfully comparable with experiment, consistent with the experimental results for carbopol and mayonnaise in Figs.~\ref{fig:carbopol} and~\ref{fig:mayo}. The study of Puisto et al.~\cite{Puisto:2015} did not consider viscosity bifurcating fluids, to which we now turn our attention.


\section{Results: viscosity bifurcating yield stress fluids}
\label{sec:vb}

\subsection{Experimental}

\label{sec:exp_vb}

\begin{figure}
	\centering
	\includegraphics[width=8.0cm]{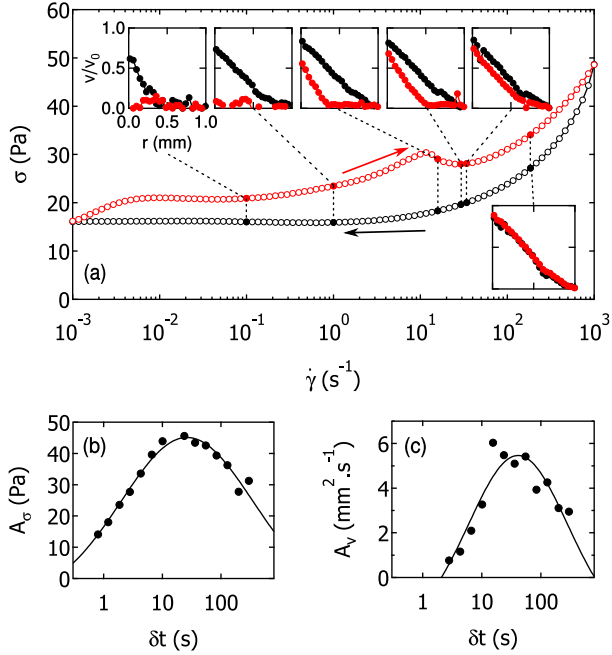}
   \caption{(a) Flow curve $\sigma$ vs. $\dot \gamma$ of a 2.5\%~wt. laponite suspension obtained by first decreasing $\dot \gamma$ from 10$^3$ to 10$^{-3}$~s$^{-1}$ in 90 logarithmically spaced steps of duration $\delta t=15.5$~s each (black symbols), and then increasing $\dot \gamma$ over the same range (red symbols). Insets: Velocity profiles inside the gap recorded at the same shear rate during the downward (black) and upward (red) sweeps. Velocity data are normalized by the rotor velocity $v_0$ and the vertical scale goes from 0 to 1. (b) Hysteresis loop area $A_{\sigma}$ defined by Eq.~\eqref{eq:Abulk} versus $\delta t$. (c) Area $A_v$ defined from the velocity profiles by Eq.~\eqref{eq:Aprofile} versus $\delta t$. The bell-shaped solid lines in (b) and (c) show the existence of a maximum in both $A_\sigma$ and $A_v$. Experiments conducted with rough boundary conditions (sandblasted Plexiglas).}
\label{fig:laponite}
 \end{figure}

Fig.~\ref{fig:laponite} shows the experimental results obtained on a 2.5\%~wt. laponite suspension in rough boundary conditions (sandblasted Plexiglas) for $\delta t=15.5$~s. On the downsweep, the flow remains homogeneous down to $\dot \gamma \simeq 1$~s$^{-1}$. Below that point, the stress reaches an almost flat plateau and shear gets localised close to the rotor while the material is arrested close to the stator. Such shear localisation is expected for viscosity bifurcating yield stress fluids and has been observed many times in steady-states obtained from fully fluidized states at large shear rates.~\cite{Coussot:2002a,Ragouilliaux:2006,Moller:2008,Martin:2012,Kurokawa:2015} 
At very low shear rates, the flow is fully arrested with total slip at the rotor. Starting up again from this arrested state, total wall-slip persists up to a stress maximum for $\dot \gamma \simeq 10$~s$^{-1}$ after which a fluidized shear-band grows from the rotor
 while the stress decreases for $10\lesssim \dot \gamma \lesssim 30$~s$^{-1}$. Along the final increasing branch of the upsweep, the laponite suspension is fully fluidized and the flow is homogeneous.
 
 In contrast to simple yield stress fluids, the hysteresis in Fig.~\ref{fig:laponite}(a) is very large and extends almost over the full range of shear rates. Even more strikingly, both $A_\sigma$ and $A_v$ now go through a clear maximum as a function of $\delta t$, for $\delta t \simeq 27$~s and 42~s respectively [Fig.~\ref{fig:laponite}(b,c)]. 

\begin{figure}
	\centering
	\includegraphics[width=8.0cm]{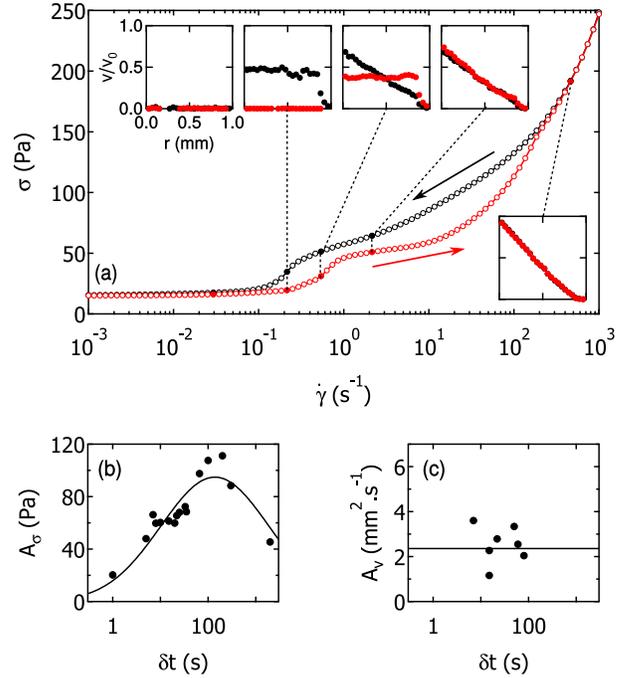}
	\caption{Same as Fig.~\ref{fig:laponite} for an 8\%~wt. carbon black gel with $\delta t=22.5$~s. Note that the stress during the downsweep (black symbols) lies above that measured during the upweep (red symbols). Experiments conducted with smooth boundary conditions (polished Plexiglas).} 
\label{fig:CB}
\end{figure}

 The same trend is observed in $A_\sigma$ for an 8\%~wt. carbon black gel in Fig.~\ref{fig:CB}(b). There again, the magnitude of the hysteresis first increases with the sweep time $\delta t$ before reaching a maximum and decreasing for the slowest sweeps. However, the flow curves shown in Fig.~\ref{fig:CB}(a) for carbon black with $\delta t=22.5$~s and in a smooth geometry are markedly different from those in Fig.~\ref{fig:laponite}(a). Here, the upsweep lies above the downsweep. This is a typical feature of rheopectic fluids where the viscosity tends to increase with time under an applied shear flow and indeed, carbon black gels have recently been recognized as rheopectic materials.~\cite{Ovarlez:2013} Since the viscosity bifurcation in yield stress fluids is associated with thixotropy, i.e. the decrease of viscosity with time under an applied shear, carbon black gels should not strickly speaking be considered here as ``viscosity bifurcating.''  Still, it is interesting to note the similarity of the behaviours of the hysteresis areas in the carbon black gel and in the laponite suspension. In particular, taking the absolute value of the stress difference in Eq.~\eqref{eq:Abulk} allows us to consider thixotropic and rheopectic materials in the same framework.  

However for carbon black gels, the loop area $A_v$ computed from the local velocity profiles strongly differs from the area $A_\sigma$ computed from the macroscopic rheology in that $A_v$ does not show any specific trend with the sweep time $\dt$. A tentative explanation lies in the complex flow behaviour of carbon black gel, which fluidisation was shown to involve flow heterogeneities not only along the radial direction but also along the vorticity direction.~\cite{Perge:2014b,Gibaud:2016} Hence one should not expect the velocity profile measured at a given height of the Couette cell to be representative of the whole sample. Furthermore, velocity profiles reported in Fig.~\ref{fig:CB}(a) show plug-like flows and much more significant wall-slip over a larger range of shear rates than for the experiments conducted on Laponite suspensions. This could also account partially for the discrepancy between the local and global areas. In any case, the fluidity model used in this paper does not capture the local behaviour of rheopectic fluids and comparisons between the model and experiments on carbon black should be taken with caution.

\begin{figure*}[!t]
	\centering
  \includegraphics[width=16cm]{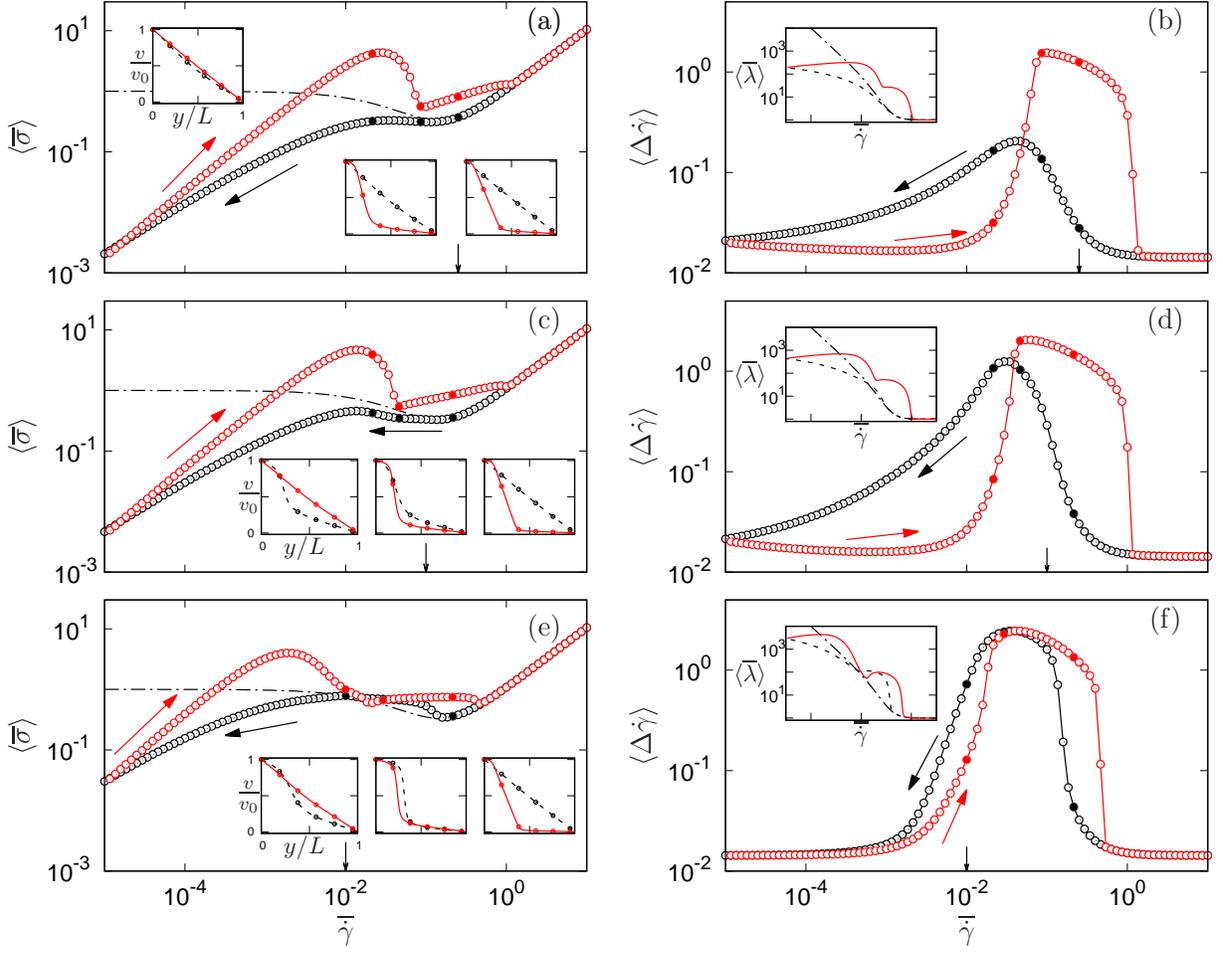}
  \caption{Viscosity bifurcating yield stress fluid model in the nearly inelastic regime, $G_0=100$, in flow curve sweeps with waiting time $\dt=3.98$ per point (top), $\dt=10$ (middle) and $\dt=100.0$ per point (bottom). In each case the left panel shows the flow curves obtained in a downsweep (black circles) followed up an upsweep (red circles), with the stationary underlying constitutive curve shown as the dot-dashed line, and the insets showing velocity profile snapshots at the times corresponding to the solid symbols in the main graph.  The right panel shows the corresponding degree of shear-banding (main graph) and the inverse fluidity (inset). The arrow pointing at the x-axis is $\ogdot=1/\dt$. Videos of the rheological hysteresis in this model for $\dt=$ 3.98, 10, and 100 can be found in the ESI.\ref{foot1}}
\label{fig:nl_100}
\end{figure*}

Whatever the mechanism for time-dependence, Figs.~\ref{fig:laponite} and \ref{fig:CB} show that both thixotropic and rheopectic yield stress fluids display similar bell-shaped loop areas for $A_\sigma$. It was proposed in Ref.~\cite{Divoux:2013} that such maxima in $A_\sigma$ (and in $A_v$ for Laponite suspensions) define a time $\delta t^\star$ that is characteristic of the material's time-dependent behaviour. For fast sweeps, $\delta t<\delta t^\star$, the material structure only partially rebuilds during the loop and the slower the sweeps, the material is given more and more time to rebuild, hence the increasing hysteresis with increasing $\delta t$. For slow sweeps, however, as $\delta t$ increases, the material reaches states that are closer and closer to steady-states at every step in shear rate. Therefore, the hysteresis magnitude shows a decreasing trend for $\delta t>\delta t^\star$. In the following, we test this intuition by studying the fluidity model in the case of a viscosity bifurcating yield stress fluid. In fact, our results will suggest that the formation of shear-bands with layer normals in the gradient direction plays an integral role in explaining the bell shaped hysteresis area as a function of time. Note that our theoretical work does not, however, take into account any flow heterogeneity along the vorticity direction.

\subsection{Theory}
\label{sec:th_vb}

\subsubsection{Fluidity model}

\begin{figure*}[!t]
	\centering
  \includegraphics[width=16cm]{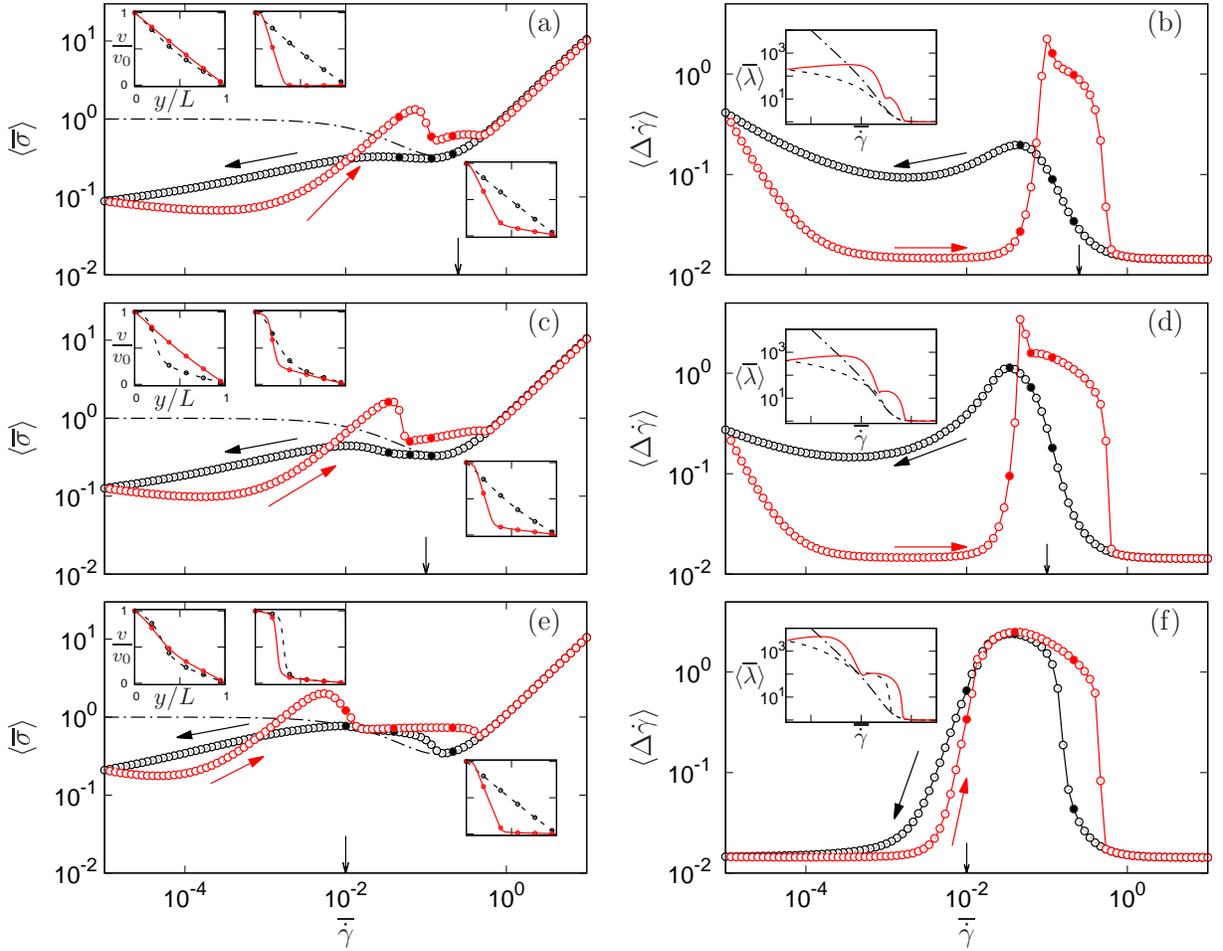}
  \caption{Viscosity bifurcating yield stress fluid model with high degree of viscoelasticity, $G_0=1$, in flow curve sweeps with waiting time $\dt=3.98$ per point (top), $\dt=10$ (middle) and $\dt=100.0$ per point (bottom). In each case the left panel shows the flow curves obtained in a downsweep (black circles) followed up an upsweep (red circles), with the stationary underlying constitutive curve shown as the dot-dashed line, and the insets showing velocity profile snapshots at the times corresponding to the solid symbols in the main graph.  The right panel shows the corresponding degree of shear-banding (main graph) and the inverse fluidity (inset). The arrow pointing at the x-axis is $\ogdot=1/\dt$. Videos of the rheological hysteresis in this model for $\dt=$ 3.98, 10, and 100 can be found in the ESI.\ref{foot1}}
\label{fig:nl_1}
\end{figure*}

The rheological hysteresis of a viscosity bifurcating yield stress fluid can be studied using a fluidity model with a non-monotonic underlying constitutive curve.  To specify a constitutive curve with a significant region of non-monotonicity, as in Fig.~\ref{fig:constit_sketch}, we set values $a=10$, $p=1$ and $\tau_1=3$ in Eq.~\eqref{eq:f_vb}.  As in our study of simple yield stress fluids above, we set the kinetic coefficient prescribing the rate at which the fluidity responds to a given flow in Eq.~\ref{eq:phi_m} as $r=1$. With this, we then consider separately the case of an almost inelastic material, with $G_0=100$ in Eq.~\ref{eq:sig_m}, and a viscoelastic fluid with $G_0=1$.  

Results for the almost inelastic case are shown in Fig.~\ref{fig:nl_100}.  As can be seen, the response shares several features in common with the counterpart results for an almost inelastic simple yield stress fluid in Fig.~\ref{fig:k_100}.  In particular, (i) at the highest strain rates the stress responses during the up and down sweeps coincide with each other, and with the underlying constitutive curve; (ii) during the downsweep the inverse fluidity $\lambda$ and stress deviate downwards from values prescribed by the constitutive curve once the shear rate becomes lower than $\ogdot \sim 1/\delta t$; (iii) the stress during the upsweep lies above that on the downsweep even at the lowest strain rates, and (iv) on the upsweep a stress overshoot is seen, closely followed by the formation of shear-bands. These bands are triggered as a result of this stress overshoot, in direct counterpart to the formation of shear-bands in shear startup from rest.~\cite{Moorcroft:2011}

However some important differences between the viscosity bifurcating results of Fig.~\ref{fig:nl_100} and the simple yield stress results of Fig.~\ref{fig:k_100} are also evident, stemming from the fact that the viscosity bifurcating fluid has a non-monotonic underlying constitutive curve with a negative slope for shear rates less than a critical rate $\gdot^*=0.332$ (recall Fig.~\ref{fig:constit_sketch}), rendering stationary homogeneous flow unstable to the formation of shear-bands in that regime. We shall now explore these differences in detail in the context separately of fast sweeps, $\delta t \lesssim 10$, and slow sweeps, $\delta t \gtrsim 10$.  

For the fast sweep shown in Figs.~\ref{fig:nl_100}(a,b), the underlying constitutive curve plays relatively little role in the response of the fluid during the downsweep, which is accordingly very similar to its counterpart for the simple yield stress fluid in Fig.~\ref{fig:k_100} (a,b). In particular, the stress already deviates downwards from the constitutive curve due to the time-dependence of the flow by the time the minimum of the constitutive curve is attained.  For strain rates lower than this, the stress and inverse fluidity fail to attain their steady-state values before the strain rate again tracks on downward.  There is therefore insufficient time for shear-bands to form, and the non-monotonicity of the underlying constitutive curve is of little consequence.

\begin{figure}[!h]
	\centering
  \includegraphics[width=8.0cm]{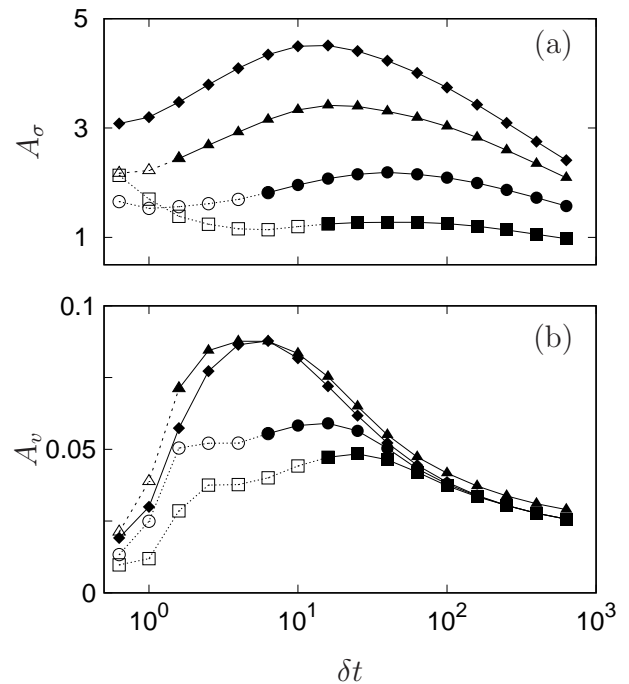}
  \caption{Area of hysteresis loops in a viscosity bifurcating yield stress fluid model as a function of sweep time per point $\dt$, as measured in (a) bulk flow curve measurements and (b) spatially resolved velocimetry. Curves upward in each panel correspond to decreasing levels of viscoelasticity: $G_0=1.0$ ($\square$), $G_0=3.16$ ($\circ$), $G_0=10.0$ ($\triangle$), and $G_0=100.0$ ($\diamondsuit$). Data points corresponding to closed hysteresis loops for which the flow curves meet at $\gdotmax$ are denoted by filled symbols; those corresponding to open loops are shown as empty symbols.}
  \label{fig:area_a1}
\end{figure}

During the fast upsweep in Figs.~\ref{fig:nl_100}(a,b), a stress overshoot arises with band formation shortly after it, as in a simple yield stress fluid. As noted above, this is the direct counterpart of the transient shear-banding triggered by stress overshoot in shear startup from rest.~\cite{Moorcroft:2011} The strain rate at which this overshoot occurs increases with decreasing sweep time $\delta t$.  In consequence, for the smallest values of $\delta t$ the stress overshoot occurs at shear rates for which the negative slope in the underlying constitutive curve is rather pronounced. This underlying instability appears to have the effect of hastening the stress drop off the overshoot, and therefore of slightly lowering the height of the overshoot.  In comparison, for slightly larger values $\delta t$, but still in the fast sweep regime $\delta t\lesssim 10$, the overshoot occurs at slightly lower values of $\ogdot$ for which the underlying negative constitutive slope is less pronounced.  This allows a higher overshoot before the stress drops away. [A careful comparison of Figs.~\ref{fig:nl_100}(a) and (c) on a linear vertical scale indeed reveals the stress overshoot to be lower in a) than (c).] This increase in the stress overshoot with $\delta t$ renders the area $\Abulk$ of the hysteresis loop an increasing function of $\delta t$ in this fast sweep regime $\delta t \lesssim 10$ (Fig.~\ref{fig:area_a1}).

In the regime of slow sweeps, $\delta t>10.0$, the response of a viscosity bifurcating fluid differs strongly from the counterpart response of a simple yield stress fluid even during the downsweep.  This can be seen by comparing Fig.~\ref{fig:nl_100}(e) with Fig.~\ref{fig:k_100}(c). In particular, in a slow downsweep in a viscosity bifurcating fluid, shear-bands have sufficient time to form in response to the negative slope in the stationary underlying constitutive curve.  At the lowest strain rates, however, these bands heal back to homogeneous flow. This is because the width of the high shear-band eventually becomes comparable to the width of the interface between the bands at low strain rates, as set by the parameter $d$ in Eq.~\ref{eq:phi_m}, thereby effectively eliminating the high shear-band. Repeating the runs in Fig.~\ref{fig:nl_100}(e) for progressively smaller values of $d$, the strain rate at which the bands heal back to homogeneous flow on the downsweep becomes progressively smaller.  We expect the value of $d$ in any experimental system to be much smaller than the smallest value that can be accessed in our simulation, and therefore that this effect would be deferred to much smaller strain rates in any real experiment. 
In the upward part of a slow sweep a stress overshoot again arises, as for fast upsweeps, triggering the formation of bands by analogy with the band formation that occurs due to stress overshoot in shear startup.  Once formed, these bands are closely analogous in this slow sweep to those seen in a steadily imposed shear flow, and persist until the strain rate exceeds that of the steady-state shear-banding regime.  The stress at which these quasi steady-state bands coexist is determined by the diffusion term in Eq.~\eqref{eq:phi_m}.  

In these slow sweeps, the height of the stress overshoot during the upsweep decreases with increasing $\delta t$, and the typical strain rate below which the response deviates from steady-state shifts to the left. In consequence, the area of the hysteresis loop decreases with increasing $\delta t$ in this slow sweep regime. See Fig.~\ref{fig:area_a1}.

The trends just discussed in an almost inelastic viscosity bifurcating fluid for the hysteresis loop area to increase with increasing $\delta t$ in fast sweeps, and to decrease with increasing $\delta t$ in slow sweeps, lead to the bell shaped dependence of $\Abulk$ on $\delta t$ in Fig.~\ref{fig:area_a1} for $G_0=100$. Experimentally, varying the  number $n$ of measurement points per decade in the shear rate sweeps reveals that the true control parameter is $n \delta t$ [Figs.~\ref{fig:several}(a) and (c)]. The same trend is also seen in the numerical results,  especially in the velocity profile hysteresis loop area $\Aprofile(\delta t)$.  Figs.~\ref{fig:several}(b) and (d) collect these curves of $\Abulk$ and $\Aprofile$ for several different values of the number $n$ of sweep points per decade, and show good collapse as a function of the master scaling variable $n\delta t$. The numerical results here allow to extend to the area computed from the velocity profile the validity of such master scaling variable.

In Fig.~\ref{fig:nl_1} we show our results for a viscosity bifurcating fluid with a high degree of viscoelasticity, $G_0=1$. The differences with the almost inelastic case of $G_0=100$ just discussed in Fig.~\ref{fig:nl_100} closely mirror the differences between a viscoelastic and almost inelastic simple yield stress fluid in Figs.~\ref{fig:k_100} and~\ref{fig:k_1} respectively.  In particular, for a viscoelastic fluid the stress evolves on the same slow timescale as the fluidity, and takes longer to respond to changes in the imposed strain rate. This leads to the upsweep flow curve initially falling below the downsweep flow curve in Fig.~\ref{fig:nl_1}(e), as in Fig.~\ref{fig:k_1}(c).

\subsubsection{Comparison with experiment and other theoretical works}

\begin{figure}
  \includegraphics[width=8.0cm]{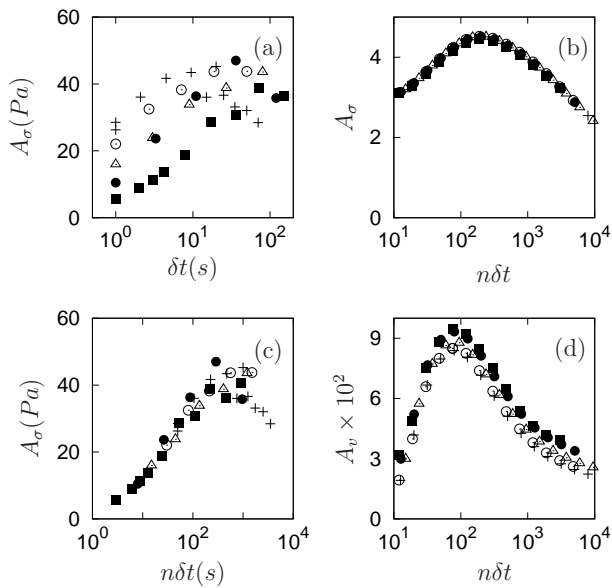}
  \caption{Area of hysteresis loops measured from bulk flow curve (a, b, c) and spatially resolved velocimetry (d) for different number of steps per decade $n=3 (\blacksquare), 8 (\bullet), 15 (\triangle), 30 (\circ), 50 (+)$ from the laponite suspension experiments (a, c), and from the viscosity bifurcating fluid model in the nearly inelastic regime $G_0=100$ (b, d). \label{fig:several}}
\end{figure}

Our fluidity model of a viscosity bifurcating yield stress fluid has successfully captured shear-banding in the downward part of a flow curve sweep [recall, {\it e.g.}, the inset of Fig.~\ref{fig:nl_100}(c)], as seen experimentally for laponite in Fig.~\ref{fig:laponite}.  This represents an important difference between the response of viscosity bifurcating and simple yield stress fluids: shear localisation was not observed during the downsweep either experimentally in simple yield stress fluids, or in our model of a simple yield stress fluid in Sec.~\ref{sec:syf}.

For a viscosity bifurcating fluid we have also successfully obtained a bell shaped dependence of the hysteresis loop areas in stress $\Abulk$ and velocities $\Aprofile$ as a function of $\delta t$, as seen in the experimental data of Sec.~\ref{sec:exp_vb}.  We also find convincing collapse of the data for $\Abulk$ and $\Aprofile$ with the scaling variable $n\delta t$ for different values of $n$ in Fig.~\ref{fig:several}(b,d), as seen experimentally in panel (c) of the same figure.  However, there is a clear separation in the location of the peak in $\Abulk(n \dt)$ compared with that for $\Aprofile(n \dt)$ in the theoretical results.  Such a separation was also noted in the previous theoretical work of Puisto et al.~\cite{Puisto:2015} for a simple yield stress fluid, although we suggest that the peak reported in that work should be treated with caution for the reasons discussed above.

\section{Conclusions}

\label{sec:conclusions}

Motivated by experimental data for two simple yield stress fluids (carbopol gel and mayonnaise) and for a viscosity bifurcating fluid (laponite), we have studied theoretically the phenomenon of rheological hysteresis in the flow curve sweep protocols as defined in the experimental study of Ref.~\cite{Divoux:2013}. We have done so within simplified fluidity models of simple yield stress and viscosity bifurcating fluid behaviour, and within the soft glassy rheology model of a simple yield stress fluid.

For a simple yield stress fluid, our simulations capture homogeneous flow response during the downsweep in strain rate, followed by shear-banding triggered by stress overshoot in the upsweep. Both these findings agree with the experimental data in carbopol. We also find that the hysteresis loop areas measured separately via the bulk stress $\Abulk$ and the spatially resolved velocity profiles $\Aprofile$ both decrease monotonically with the timescale $\delta t$ of the imposed sweep, as also seen experimentally (at least for all but the lowest value of $\delta t$ in mayonnaise).  While this comparison seems very convincing, a note of caution should also be added that, experimentally, the smallest feasible value of $\delta t$ may simply be still too large to pick up any regime in which $\Abulk$ and $\Aprofile$ rise with $\delta t$.

Our prediction of a monotonic decrease in the hysteresis loop areas with increasing $\delta t$ in a simple yield stress fluid is in contrast with the earlier theoretical work Puisto et al.~\cite{Puisto:2015}. We have argued that the non-monotonicity suggested in that earlier study stems from their having considered sweep times fast enough to be in the physically undesirable regime in which the downsweep and upsweep flow curves fail to meet at the highest strain rates, which was excluded from the experiments.

For a viscosity bifurcating yield stress fluid, our simulations capture shear-banding during slow downsweeps, arising from the instability present in the non-monotonic underlying stationary constitutive curve. This agrees with experimental data in laponite.  For fast downsweeps, in contrast, these bands do not have sufficient time to form. During the upsweep, bands form due to the overshoot of stress as a function of time, in close analogy with overshoot-driven shear-banding in shear startup from rest. As a consequence of this more complicated banding behaviour in a viscosity bifurcating yield stress fluid, we have found the hysteresis loop areas $\Abulk$ and $\Aprofile$ to show a bell shaped dependence on the sweep time $\delta t$, as also seen experimentally in laponite.

For both simple and viscosity bifurcating yield stress fluids, an important difference arises between the case of an almost inelastic material, in which the stress responds essentially instantaneously to a given imposed flow and state of material fluidity, and a viscoelastic fluid, in which the stress responds on the same slow timescale as the fluidity itself.  In particular, for almost inelastic fluids the stress during the upsweep lies above that of the downsweep for all strain rates. In contrast, for a viscoelastic fluid the stress in the first part of the upsweep lies below that in the downsweep due to the continuing slow viscoelastic relaxation of the stress. The experimental data in carbopol, mayonnaise and laponite suggest a mostly inelastic response, with a slight viscoelastic effect in carbopol at low strain rates. The carbon black gel shows a much more pronounced viscoelastic response, although with rheopexy that leads to a discrepancy between the hysteresis area computed from the velocity profiles and that computed from the rheology alone, which the theoretical models considered here do not address. Flow curve sweeps might indeed provide a good way in which to probe the degree of viscoelasticity in a soft glassy material.

A significant advantage of the fluidity model is the facility to tune between the inelastic and viscoelastic response just discussed via the parameter $G_0$.  In contrast the SGR model predicts only viscoelastic behaviour, with the stress responding on the same timescale as the material's fluidity.  Along with the relatively much greater simplicity of the fluidity model compared to SGR, this suggests an advantage of fluidity models over SGR in modelling the experimental phenomena considered here.  Note that we have considered in this study only a simple yield stress fluid version of the SGR model. A counterpart version capturing (similarly viscoelastic) viscosity bifurcating behaviour could also have been studied.

In all our theoretical calculations we have ignored wall-slip, assuming always a no-slip condition at the walls of the flow cell. Given the strong presence of slip in many of the experiments, this represents a major shortcoming that remains to be tackled in future theoretical work.

Finally, the present study has for simplicity taken a scalar approach, considering only the shear component of the stress tensor.  In practice, other components of the stress tensor are also important to the rheology of soft glassy materials, particularly to model some three-dimensional features observed during yielding,~\cite{Perge:2014b,Gibaud:2016} or under confinement.~\cite{Montesi:2004,LinGibson:2004,Osuji2008,Grenard:2011} These can be incorporated by using a frame invariant tensorial constitutive model, such as the Maxwell or Giesekus model, in which the material viscosity depends on the scalar fluidity $\lambda$.~\cite{Mewis:2009,deSouzaMendes:2012,Armstrong:2016} A tensorial extension of the scalar SGR model can be found in Ref.~\cite{Cates:2004}. A study of rheohysteresis in these tensorial models remains open for future work.


\section*{Acknowledgements}
The authors thank Dr. Ricard Matas Navarro, Prof. Catherine Barentin and Dr. Vincent Grenard for helpful discussions and technical help.  SMF's and RR's research leading to these results has received funding from the European Research Council under the European Union's Seventh Framework Programme (FP7/2007-2013)/ERC grant agreement number 279365.  SM and TD acknowledge funding from the European Research Council under the European Union's Seventh Framework Programme (FP7/2007-2013)/ERC grant agreement number 258803. 



\bibliography{new_article} 

\begin{thebibliography}{93}%
\makeatletter
\providecommand \@ifxundefined [1]{%
 \@ifx{#1\undefined}
}%
\providecommand \@ifnum [1]{%
 \ifnum #1\expandafter \@firstoftwo
 \else \expandafter \@secondoftwo
 \fi
}%
\providecommand \@ifx [1]{%
 \ifx #1\expandafter \@firstoftwo
 \else \expandafter \@secondoftwo
 \fi
}%
\providecommand \natexlab [1]{#1}%
\providecommand \enquote  [1]{``#1''}%
\providecommand \bibnamefont  [1]{#1}%
\providecommand \bibfnamefont [1]{#1}%
\providecommand \citenamefont [1]{#1}%
\providecommand \href@noop [0]{\@secondoftwo}%
\providecommand \href [0]{\begingroup \@sanitize@url \@href}%
\providecommand \@href[1]{\@@startlink{#1}\@@href}%
\providecommand \@@href[1]{\endgroup#1\@@endlink}%
\providecommand \@sanitize@url [0]{\catcode `\\12\catcode `\$12\catcode
  `\&12\catcode `\#12\catcode `\^12\catcode `\_12\catcode `\%12\relax}%
\providecommand \@@startlink[1]{}%
\providecommand \@@endlink[0]{}%
\providecommand \url  [0]{\begingroup\@sanitize@url \@url }%
\providecommand \@url [1]{\endgroup\@href {#1}{\urlprefix }}%
\providecommand \urlprefix  [0]{URL }%
\providecommand \Eprint [0]{\href }%
\providecommand \doibase [0]{http://dx.doi.org/}%
\providecommand \selectlanguage [0]{\@gobble}%
\providecommand \bibinfo  [0]{\@secondoftwo}%
\providecommand \bibfield  [0]{\@secondoftwo}%
\providecommand \translation [1]{[#1]}%
\providecommand \BibitemOpen [0]{}%
\providecommand \bibitemStop [0]{}%
\providecommand \bibitemNoStop [0]{.\EOS\space}%
\providecommand \EOS [0]{\spacefactor3000\relax}%
\providecommand \BibitemShut  [1]{\csname bibitem#1\endcsname}%
\let\auto@bib@innerbib\@empty
\bibitem [{\citenamefont {B{\'e}cu}\ \emph {et~al.}(2006)\citenamefont
  {B{\'e}cu}, \citenamefont {Manneville},\ and\ \citenamefont
  {Colin}}]{Becu:2006}%
  \BibitemOpen
  \bibfield  {author} {\bibinfo {author} {\bibfnamefont {L.}~\bibnamefont
  {B{\'e}cu}}, \bibinfo {author} {\bibfnamefont {S.}~\bibnamefont
  {Manneville}}, \ and\ \bibinfo {author} {\bibfnamefont {A.}~\bibnamefont
  {Colin}},\ }\href@noop {} {\bibfield  {journal} {\bibinfo  {journal} {Phys.
  Rev. Lett.}\ }\textbf {\bibinfo {volume} {96}},\ \bibinfo {pages} {138302}
  (\bibinfo {year} {2006})}\BibitemShut {NoStop}%
\bibitem [{\citenamefont {Rouyer}\ \emph {et~al.}(2008)\citenamefont {Rouyer},
  \citenamefont {Cohen-Addad}, \citenamefont {H{\"{o}}hler}, \citenamefont
  {Sollich},\ and\ \citenamefont {Fielding}}]{Rouyer2008}%
  \BibitemOpen
  \bibfield  {author} {\bibinfo {author} {\bibfnamefont {F.}~\bibnamefont
  {Rouyer}}, \bibinfo {author} {\bibfnamefont {S.}~\bibnamefont {Cohen-Addad}},
  \bibinfo {author} {\bibfnamefont {R.}~\bibnamefont {H{\"{o}}hler}}, \bibinfo
  {author} {\bibfnamefont {P.}~\bibnamefont {Sollich}}, \ and\ \bibinfo
  {author} {\bibfnamefont {S.~M.}\ \bibnamefont {Fielding}},\ }\href {\doibase
  10.1140/epje/i2008-10382-7} {\bibfield  {journal} {\bibinfo  {journal} {Eur.
  Phys. J. E}\ }\textbf {\bibinfo {volume} {27}},\ \bibinfo {pages} {309}
  (\bibinfo {year} {2008})}\BibitemShut {NoStop}%
\bibitem [{\citenamefont {Mason}\ and\ \citenamefont
  {Weitz}(1995)}]{Mason:1995b}%
  \BibitemOpen
  \bibfield  {author} {\bibinfo {author} {\bibfnamefont {T.~G.}\ \bibnamefont
  {Mason}}\ and\ \bibinfo {author} {\bibfnamefont {D.~A.}\ \bibnamefont
  {Weitz}},\ }\href@noop {} {\bibfield  {journal} {\bibinfo  {journal} {Phys.
  Rev. Lett.}\ }\textbf {\bibinfo {volume} {75}},\ \bibinfo {pages} {2770}
  (\bibinfo {year} {1995})}\BibitemShut {NoStop}%
\bibitem [{\citenamefont {Knaebel}\ \emph {et~al.}(2000)\citenamefont
  {Knaebel}, \citenamefont {Bellour}, \citenamefont {Munch}, \citenamefont
  {Viasnoff}, \citenamefont {Lequeux},\ and\ \citenamefont
  {Harden}}]{Knaebel:2000}%
  \BibitemOpen
  \bibfield  {author} {\bibinfo {author} {\bibfnamefont {A.}~\bibnamefont
  {Knaebel}}, \bibinfo {author} {\bibfnamefont {M.}~\bibnamefont {Bellour}},
  \bibinfo {author} {\bibfnamefont {J.-P.}\ \bibnamefont {Munch}}, \bibinfo
  {author} {\bibfnamefont {V.}~\bibnamefont {Viasnoff}}, \bibinfo {author}
  {\bibfnamefont {F.}~\bibnamefont {Lequeux}}, \ and\ \bibinfo {author}
  {\bibfnamefont {J.~L.}\ \bibnamefont {Harden}},\ }\href@noop {} {\bibfield
  {journal} {\bibinfo  {journal} {Europhys. Lett.}\ }\textbf {\bibinfo {volume}
  {52}},\ \bibinfo {pages} {73} (\bibinfo {year} {2000})}\BibitemShut {NoStop}%
\bibitem [{\citenamefont {Cloitre}\ \emph {et~al.}(2000)\citenamefont
  {Cloitre}, \citenamefont {Borrega},\ and\ \citenamefont
  {Leibler}}]{Cloitre:2000}%
  \BibitemOpen
  \bibfield  {author} {\bibinfo {author} {\bibfnamefont {M.}~\bibnamefont
  {Cloitre}}, \bibinfo {author} {\bibfnamefont {R.}~\bibnamefont {Borrega}}, \
  and\ \bibinfo {author} {\bibfnamefont {L.}~\bibnamefont {Leibler}},\
  }\href@noop {} {\bibfield  {journal} {\bibinfo  {journal} {Phys. Rev. Lett.}\
  }\textbf {\bibinfo {volume} {85}},\ \bibinfo {pages} {4819} (\bibinfo {year}
  {2000})}\BibitemShut {NoStop}%
\bibitem [{\citenamefont {Rogers}\ \emph {et~al.}(2010)\citenamefont {Rogers},
  \citenamefont {Callaghan}, \citenamefont {Petekidis},\ and\ \citenamefont
  {Vlassopoulos}}]{Rogers:2010}%
  \BibitemOpen
  \bibfield  {author} {\bibinfo {author} {\bibfnamefont {S.}~\bibnamefont
  {Rogers}}, \bibinfo {author} {\bibfnamefont {P.}~\bibnamefont {Callaghan}},
  \bibinfo {author} {\bibfnamefont {G.}~\bibnamefont {Petekidis}}, \ and\
  \bibinfo {author} {\bibfnamefont {D.}~\bibnamefont {Vlassopoulos}},\
  }\href@noop {} {\bibfield  {journal} {\bibinfo  {journal} {J. Rheol.}\
  }\textbf {\bibinfo {volume} {54}},\ \bibinfo {pages} {133} (\bibinfo {year}
  {2010})}\BibitemShut {NoStop}%
\bibitem [{\citenamefont {Bonnecaze}\ and\ \citenamefont
  {Cloitre}(2010)}]{Bonnecaze:2010}%
  \BibitemOpen
  \bibfield  {author} {\bibinfo {author} {\bibfnamefont {R.}~\bibnamefont
  {Bonnecaze}}\ and\ \bibinfo {author} {\bibfnamefont {M.}~\bibnamefont
  {Cloitre}},\ }\href@noop {} {\bibfield  {journal} {\bibinfo  {journal} {Adv.
  Polym. Sci.}\ }\textbf {\bibinfo {volume} {236}},\ \bibinfo {pages} {117}
  (\bibinfo {year} {2010})}\BibitemShut {NoStop}%
\bibitem [{\citenamefont {Bonn}\ \emph {et~al.}(2015)\citenamefont {Bonn},
  \citenamefont {Paredes}, \citenamefont {Denn}, \citenamefont {Berthier},
  \citenamefont {Divoux},\ and\ \citenamefont {Manneville}}]{Bonn:2015}%
  \BibitemOpen
  \bibfield  {author} {\bibinfo {author} {\bibfnamefont {D.}~\bibnamefont
  {Bonn}}, \bibinfo {author} {\bibfnamefont {J.}~\bibnamefont {Paredes}},
  \bibinfo {author} {\bibfnamefont {M.}~\bibnamefont {Denn}}, \bibinfo {author}
  {\bibfnamefont {L.}~\bibnamefont {Berthier}}, \bibinfo {author}
  {\bibfnamefont {T.}~\bibnamefont {Divoux}}, \ and\ \bibinfo {author}
  {\bibfnamefont {S.}~\bibnamefont {Manneville}},\ }\href@noop {} {\bibfield
  {journal} {\bibinfo  {journal} {arXiv:1502.05281}\ } (\bibinfo {year}
  {2015})}\BibitemShut {NoStop}%
\bibitem [{\citenamefont {Fielding}\ \emph {et~al.}(2000)\citenamefont
  {Fielding}, \citenamefont {Sollich},\ and\ \citenamefont
  {Cates}}]{Fielding:2000}%
  \BibitemOpen
  \bibfield  {author} {\bibinfo {author} {\bibfnamefont {S.~M.}\ \bibnamefont
  {Fielding}}, \bibinfo {author} {\bibfnamefont {P.}~\bibnamefont {Sollich}}, \
  and\ \bibinfo {author} {\bibfnamefont {M.~E.}\ \bibnamefont {Cates}},\ }\href
  {\doibase 10.1122/1.551088} {\bibfield  {journal} {\bibinfo  {journal} {J.
  Rheol.}\ }\textbf {\bibinfo {volume} {44}},\ \bibinfo {pages} {323} (\bibinfo
  {year} {2000})}\BibitemShut {NoStop}%
\bibitem [{\citenamefont {Cloitre}\ \emph {et~al.}(2003)\citenamefont
  {Cloitre}, \citenamefont {Borrega}, \citenamefont {Monti},\ and\
  \citenamefont {Leibler}}]{Cloitre:2003}%
  \BibitemOpen
  \bibfield  {author} {\bibinfo {author} {\bibfnamefont {M.}~\bibnamefont
  {Cloitre}}, \bibinfo {author} {\bibfnamefont {R.}~\bibnamefont {Borrega}},
  \bibinfo {author} {\bibfnamefont {F.}~\bibnamefont {Monti}}, \ and\ \bibinfo
  {author} {\bibfnamefont {L.}~\bibnamefont {Leibler}},\ }\href@noop {}
  {\bibfield  {journal} {\bibinfo  {journal} {Phys. Rev. Lett.}\ }\textbf
  {\bibinfo {volume} {90}},\ \bibinfo {pages} {068303} (\bibinfo {year}
  {2003})}\BibitemShut {NoStop}%
\bibitem [{\citenamefont {Negi}\ and\ \citenamefont
  {Osuji}(2010)}]{Negi:2010b}%
  \BibitemOpen
  \bibfield  {author} {\bibinfo {author} {\bibfnamefont {A.}~\bibnamefont
  {Negi}}\ and\ \bibinfo {author} {\bibfnamefont {C.}~\bibnamefont {Osuji}},\
  }\href@noop {} {\bibfield  {journal} {\bibinfo  {journal} {Phys. Rev. E}\
  }\textbf {\bibinfo {volume} {82}},\ \bibinfo {pages} {031404} (\bibinfo
  {year} {2010})}\BibitemShut {NoStop}%
\bibitem [{\citenamefont {Ovarlez}\ \emph
  {et~al.}(2013{\natexlab{a}})\citenamefont {Ovarlez}, \citenamefont
  {Cohen-Addad}, \citenamefont {Krishan}, \citenamefont {Goyon},\ and\
  \citenamefont {Coussot}}]{Ovarlez:2013b}%
  \BibitemOpen
  \bibfield  {author} {\bibinfo {author} {\bibfnamefont {G.}~\bibnamefont
  {Ovarlez}}, \bibinfo {author} {\bibfnamefont {S.}~\bibnamefont
  {Cohen-Addad}}, \bibinfo {author} {\bibfnamefont {K.}~\bibnamefont
  {Krishan}}, \bibinfo {author} {\bibfnamefont {J.}~\bibnamefont {Goyon}}, \
  and\ \bibinfo {author} {\bibfnamefont {P.}~\bibnamefont {Coussot}},\
  }\href@noop {} {\bibfield  {journal} {\bibinfo  {journal} {J. Non-Newtonian
  Fluid Mech.}\ }\textbf {\bibinfo {volume} {193}},\ \bibinfo {pages} {68}
  (\bibinfo {year} {2013}{\natexlab{a}})}\BibitemShut {NoStop}%
\bibitem [{\citenamefont {Irani}\ \emph {et~al.}(2014)\citenamefont {Irani},
  \citenamefont {Chaudhuri},\ and\ \citenamefont {Heussinger}}]{Irani:2014}%
  \BibitemOpen
  \bibfield  {author} {\bibinfo {author} {\bibfnamefont {E.}~\bibnamefont
  {Irani}}, \bibinfo {author} {\bibfnamefont {P.}~\bibnamefont {Chaudhuri}}, \
  and\ \bibinfo {author} {\bibfnamefont {C.}~\bibnamefont {Heussinger}},\
  }\href@noop {} {\bibfield  {journal} {\bibinfo  {journal} {Phys. Rev. Lett.}\
  }\textbf {\bibinfo {volume} {112}},\ \bibinfo {pages} {188303} (\bibinfo
  {year} {2014})}\BibitemShut {NoStop}%
\bibitem [{\citenamefont {Ragouilliaux}\ \emph {et~al.}(2007)\citenamefont
  {Ragouilliaux}, \citenamefont {Ovarlez}, \citenamefont {Shahidzadeh-Bonn},
  \citenamefont {Herzhaft}, \citenamefont {Palermo},\ and\ \citenamefont
  {Coussot}}]{Ragouilliaux:2007}%
  \BibitemOpen
  \bibfield  {author} {\bibinfo {author} {\bibfnamefont {A.}~\bibnamefont
  {Ragouilliaux}}, \bibinfo {author} {\bibfnamefont {G.}~\bibnamefont
  {Ovarlez}}, \bibinfo {author} {\bibfnamefont {N.}~\bibnamefont
  {Shahidzadeh-Bonn}}, \bibinfo {author} {\bibfnamefont {B.}~\bibnamefont
  {Herzhaft}}, \bibinfo {author} {\bibfnamefont {T.}~\bibnamefont {Palermo}}, \
  and\ \bibinfo {author} {\bibfnamefont {P.}~\bibnamefont {Coussot}},\
  }\href@noop {} {\bibfield  {journal} {\bibinfo  {journal} {Phys. Rev. E}\
  }\textbf {\bibinfo {volume} {76}},\ \bibinfo {pages} {051408} (\bibinfo
  {year} {2007})}\BibitemShut {NoStop}%
\bibitem [{\citenamefont {Ovarlez}\ \emph {et~al.}(2009)\citenamefont
  {Ovarlez}, \citenamefont {Rodts}, \citenamefont {Chateau},\ and\
  \citenamefont {Coussot}}]{Ovarlez:2009}%
  \BibitemOpen
  \bibfield  {author} {\bibinfo {author} {\bibfnamefont {G.}~\bibnamefont
  {Ovarlez}}, \bibinfo {author} {\bibfnamefont {S.}~\bibnamefont {Rodts}},
  \bibinfo {author} {\bibfnamefont {X.}~\bibnamefont {Chateau}}, \ and\
  \bibinfo {author} {\bibfnamefont {P.}~\bibnamefont {Coussot}},\ }\href@noop
  {} {\bibfield  {journal} {\bibinfo  {journal} {Rheol. Acta}\ }\textbf
  {\bibinfo {volume} {48}},\ \bibinfo {pages} {831} (\bibinfo {year}
  {2009})}\BibitemShut {NoStop}%
\bibitem [{\citenamefont {Martin}\ and\ \citenamefont
  {Hu}(2012)}]{Martin:2012}%
  \BibitemOpen
  \bibfield  {author} {\bibinfo {author} {\bibfnamefont {J.}~\bibnamefont
  {Martin}}\ and\ \bibinfo {author} {\bibfnamefont {Y.}~\bibnamefont {Hu}},\
  }\href@noop {} {\bibfield  {journal} {\bibinfo  {journal} {Soft Matter}\
  }\textbf {\bibinfo {volume} {8}},\ \bibinfo {pages} {6940} (\bibinfo {year}
  {2012})}\BibitemShut {NoStop}%
\bibitem [{\citenamefont {Cheddadi}\ \emph {et~al.}(2012)\citenamefont
  {Cheddadi}, \citenamefont {Saramito},\ and\ \citenamefont
  {Graner}}]{Cheddadi:2012}%
  \BibitemOpen
  \bibfield  {author} {\bibinfo {author} {\bibfnamefont {I.}~\bibnamefont
  {Cheddadi}}, \bibinfo {author} {\bibfnamefont {P.}~\bibnamefont {Saramito}},
  \ and\ \bibinfo {author} {\bibfnamefont {F.}~\bibnamefont {Graner}},\
  }\href@noop {} {\bibfield  {journal} {\bibinfo  {journal} {J. Rheol.}\
  }\textbf {\bibinfo {volume} {56}},\ \bibinfo {pages} {213} (\bibinfo {year}
  {2012})}\BibitemShut {NoStop}%
\bibitem [{\citenamefont {Kurokawa}\ \emph {et~al.}(2015)\citenamefont
  {Kurokawa}, \citenamefont {Vidal}, \citenamefont {Kurita}, \citenamefont
  {Divoux},\ and\ \citenamefont {Manneville}}]{Kurokawa:2015}%
  \BibitemOpen
  \bibfield  {author} {\bibinfo {author} {\bibfnamefont {A.}~\bibnamefont
  {Kurokawa}}, \bibinfo {author} {\bibfnamefont {V.}~\bibnamefont {Vidal}},
  \bibinfo {author} {\bibfnamefont {K.}~\bibnamefont {Kurita}}, \bibinfo
  {author} {\bibfnamefont {T.}~\bibnamefont {Divoux}}, \ and\ \bibinfo {author}
  {\bibfnamefont {S.}~\bibnamefont {Manneville}},\ }\href@noop {} {\bibfield
  {journal} {\bibinfo  {journal} {Soft Matter}\ }\textbf {\bibinfo {volume}
  {11}},\ \bibinfo {pages} {9026s} (\bibinfo {year} {2015})}\BibitemShut
  {NoStop}%
\bibitem [{\citenamefont {Gibaud}\ \emph {et~al.}(2008)\citenamefont {Gibaud},
  \citenamefont {Barentin},\ and\ \citenamefont {Manneville}}]{Gibaud:2008}%
  \BibitemOpen
  \bibfield  {author} {\bibinfo {author} {\bibfnamefont {T.}~\bibnamefont
  {Gibaud}}, \bibinfo {author} {\bibfnamefont {C.}~\bibnamefont {Barentin}}, \
  and\ \bibinfo {author} {\bibfnamefont {S.}~\bibnamefont {Manneville}},\
  }\href@noop {} {\bibfield  {journal} {\bibinfo  {journal} {Phys. Rev. Lett.}\
  }\textbf {\bibinfo {volume} {101}},\ \bibinfo {pages} {258302} (\bibinfo
  {year} {2008})}\BibitemShut {NoStop}%
\bibitem [{\citenamefont {Gibaud}\ \emph {et~al.}(2009)\citenamefont {Gibaud},
  \citenamefont {Barentin}, \citenamefont {Taberlet},\ and\ \citenamefont
  {Manneville}}]{Gibaud:2009}%
  \BibitemOpen
  \bibfield  {author} {\bibinfo {author} {\bibfnamefont {T.}~\bibnamefont
  {Gibaud}}, \bibinfo {author} {\bibfnamefont {C.}~\bibnamefont {Barentin}},
  \bibinfo {author} {\bibfnamefont {N.}~\bibnamefont {Taberlet}}, \ and\
  \bibinfo {author} {\bibfnamefont {S.}~\bibnamefont {Manneville}},\
  }\href@noop {} {\bibfield  {journal} {\bibinfo  {journal} {Soft Matter}\
  }\textbf {\bibinfo {volume} {5}},\ \bibinfo {pages} {3026} (\bibinfo {year}
  {2009})}\BibitemShut {NoStop}%
\bibitem [{\citenamefont {Sollich}\ \emph {et~al.}(1997)\citenamefont
  {Sollich}, \citenamefont {Lequeux}, \citenamefont {H\'ebraud},\ and\
  \citenamefont {Cates}}]{Sollich:1997}%
  \BibitemOpen
  \bibfield  {author} {\bibinfo {author} {\bibfnamefont {P.}~\bibnamefont
  {Sollich}}, \bibinfo {author} {\bibfnamefont {F.}~\bibnamefont {Lequeux}},
  \bibinfo {author} {\bibfnamefont {P.}~\bibnamefont {H\'ebraud}}, \ and\
  \bibinfo {author} {\bibfnamefont {M.~E.}\ \bibnamefont {Cates}},\ }\href@noop
  {} {\bibfield  {journal} {\bibinfo  {journal} {Phys. Rev. Lett.}\ }\textbf
  {\bibinfo {volume} {78}},\ \bibinfo {pages} {2020} (\bibinfo {year}
  {1997})}\BibitemShut {NoStop}%
\bibitem [{\citenamefont {Sollich}(1998)}]{Sollich:1998}%
  \BibitemOpen
  \bibfield  {author} {\bibinfo {author} {\bibfnamefont {P.}~\bibnamefont
  {Sollich}},\ }\href@noop {} {\bibfield  {journal} {\bibinfo  {journal} {Phys.
  Rev. E}\ }\textbf {\bibinfo {volume} {58}},\ \bibinfo {pages} {738} (\bibinfo
  {year} {1998})}\BibitemShut {NoStop}%
\bibitem [{\citenamefont {Fielding}(2014)}]{Fielding:2014}%
  \BibitemOpen
  \bibfield  {author} {\bibinfo {author} {\bibfnamefont {S.}~\bibnamefont
  {Fielding}},\ }\href@noop {} {\bibfield  {journal} {\bibinfo  {journal} {Rep.
  Prog. Phys.}\ }\textbf {\bibinfo {volume} {77}},\ \bibinfo {pages} {102601}
  (\bibinfo {year} {2014})}\BibitemShut {NoStop}%
\bibitem [{\citenamefont {Divoux}\ \emph {et~al.}(2016)\citenamefont {Divoux},
  \citenamefont {Fardin}, \citenamefont {Manneville},\ and\ \citenamefont
  {Lerouge}}]{Divoux:2016}%
  \BibitemOpen
  \bibfield  {author} {\bibinfo {author} {\bibfnamefont {T.}~\bibnamefont
  {Divoux}}, \bibinfo {author} {\bibfnamefont {M.-A.}\ \bibnamefont {Fardin}},
  \bibinfo {author} {\bibfnamefont {S.}~\bibnamefont {Manneville}}, \ and\
  \bibinfo {author} {\bibfnamefont {S.}~\bibnamefont {Lerouge}},\ }\href@noop
  {} {\bibfield  {journal} {\bibinfo  {journal} {Annu. Rev. Fluid Mech.}\
  }\textbf {\bibinfo {volume} {48}},\ \bibinfo {pages} {81} (\bibinfo {year}
  {2016})}\BibitemShut {NoStop}%
\bibitem [{\citenamefont {Herschel}\ and\ \citenamefont
  {Bulkley}(1926)}]{Herschel:1926}%
  \BibitemOpen
  \bibfield  {author} {\bibinfo {author} {\bibfnamefont {W.}~\bibnamefont
  {Herschel}}\ and\ \bibinfo {author} {\bibfnamefont {R.}~\bibnamefont
  {Bulkley}},\ }\href@noop {} {\bibfield  {journal} {\bibinfo  {journal}
  {Kolloid Zeitschrift}\ }\textbf {\bibinfo {volume} {39}},\ \bibinfo {pages}
  {291} (\bibinfo {year} {1926})}\BibitemShut {NoStop}%
\bibitem [{\citenamefont {Barry}\ and\ \citenamefont
  {Meyer}(1979)}]{Barry:1979}%
  \BibitemOpen
  \bibfield  {author} {\bibinfo {author} {\bibfnamefont {B.}~\bibnamefont
  {Barry}}\ and\ \bibinfo {author} {\bibfnamefont {M.}~\bibnamefont {Meyer}},\
  }\href@noop {} {\bibfield  {journal} {\bibinfo  {journal} {Int. J. Pharm.}\
  }\textbf {\bibinfo {volume} {2}},\ \bibinfo {pages} {1} (\bibinfo {year}
  {1979})}\BibitemShut {NoStop}%
\bibitem [{\citenamefont {Chen}\ \emph {et~al.}(1992)\citenamefont {Chen},
  \citenamefont {Zukoski}, \citenamefont {Ackerson}, \citenamefont {Hanley},
  \citenamefont {Straty}, \citenamefont {Barker},\ and\ \citenamefont
  {Glinka}}]{Chen:1992}%
  \BibitemOpen
  \bibfield  {author} {\bibinfo {author} {\bibfnamefont {L.~B.}\ \bibnamefont
  {Chen}}, \bibinfo {author} {\bibfnamefont {C.~F.}\ \bibnamefont {Zukoski}},
  \bibinfo {author} {\bibfnamefont {B.~J.}\ \bibnamefont {Ackerson}}, \bibinfo
  {author} {\bibfnamefont {H.~J.~M.}\ \bibnamefont {Hanley}}, \bibinfo {author}
  {\bibfnamefont {G.~C.}\ \bibnamefont {Straty}}, \bibinfo {author}
  {\bibfnamefont {J.}~\bibnamefont {Barker}}, \ and\ \bibinfo {author}
  {\bibfnamefont {C.~J.}\ \bibnamefont {Glinka}},\ }\href {\doibase
  10.1103/PhysRevLett.69.688} {\bibfield  {journal} {\bibinfo  {journal} {Phys.
  Rev. Lett.}\ }\textbf {\bibinfo {volume} {69}},\ \bibinfo {pages} {688}
  (\bibinfo {year} {1992})}\BibitemShut {NoStop}%
\bibitem [{\citenamefont {Cruz}\ \emph {et~al.}(2002)\citenamefont {Cruz},
  \citenamefont {Chevoir}, \citenamefont {Bonn},\ and\ \citenamefont
  {Coussot}}]{DaCruz:2002}%
  \BibitemOpen
  \bibfield  {author} {\bibinfo {author} {\bibfnamefont {F.~D.}\ \bibnamefont
  {Cruz}}, \bibinfo {author} {\bibfnamefont {F.}~\bibnamefont {Chevoir}},
  \bibinfo {author} {\bibfnamefont {D.}~\bibnamefont {Bonn}}, \ and\ \bibinfo
  {author} {\bibfnamefont {P.}~\bibnamefont {Coussot}},\ }\href@noop {}
  {\bibfield  {journal} {\bibinfo  {journal} {Phys. Rev. E}\ }\textbf {\bibinfo
  {volume} {66}},\ \bibinfo {pages} {051305} (\bibinfo {year}
  {2002})}\BibitemShut {NoStop}%
\bibitem [{\citenamefont {Holmes}\ \emph {et~al.}(2004)\citenamefont {Holmes},
  \citenamefont {Callaghan}, \citenamefont {Vlassopoulos},\ and\ \citenamefont
  {Roovers}}]{Holmes:2004}%
  \BibitemOpen
  \bibfield  {author} {\bibinfo {author} {\bibfnamefont {W.~M.}\ \bibnamefont
  {Holmes}}, \bibinfo {author} {\bibfnamefont {P.~T.}\ \bibnamefont
  {Callaghan}}, \bibinfo {author} {\bibfnamefont {D.}~\bibnamefont
  {Vlassopoulos}}, \ and\ \bibinfo {author} {\bibfnamefont {J.}~\bibnamefont
  {Roovers}},\ }\href@noop {} {\bibfield  {journal} {\bibinfo  {journal} {J.
  Rheol.}\ }\textbf {\bibinfo {volume} {48}},\ \bibinfo {pages} {1085}
  (\bibinfo {year} {2004})}\BibitemShut {NoStop}%
\bibitem [{\citenamefont {ten Brinke}\ \emph {et~al.}(2007)\citenamefont {ten
  Brinke}, \citenamefont {Bailey}, \citenamefont {Lekkerkerker},\ and\
  \citenamefont {Maitland}}]{tenBrinke:2007}%
  \BibitemOpen
  \bibfield  {author} {\bibinfo {author} {\bibfnamefont {A.}~\bibnamefont {ten
  Brinke}}, \bibinfo {author} {\bibfnamefont {L.}~\bibnamefont {Bailey}},
  \bibinfo {author} {\bibfnamefont {H.}~\bibnamefont {Lekkerkerker}}, \ and\
  \bibinfo {author} {\bibfnamefont {G.}~\bibnamefont {Maitland}},\ }\href@noop
  {} {\bibfield  {journal} {\bibinfo  {journal} {Soft Matter}\ }\textbf
  {\bibinfo {volume} {3}},\ \bibinfo {pages} {1145} (\bibinfo {year}
  {2007})}\BibitemShut {NoStop}%
\bibitem [{\citenamefont {M{\o}ller}\ \emph {et~al.}(2009)\citenamefont
  {M{\o}ller}, \citenamefont {Fall}, \citenamefont {Chikkadi}, \citenamefont
  {Derks},\ and\ \citenamefont {Bonn}}]{Moller:2009b}%
  \BibitemOpen
  \bibfield  {author} {\bibinfo {author} {\bibfnamefont {P.~C.~F.}\
  \bibnamefont {M{\o}ller}}, \bibinfo {author} {\bibfnamefont {A.}~\bibnamefont
  {Fall}}, \bibinfo {author} {\bibfnamefont {V.}~\bibnamefont {Chikkadi}},
  \bibinfo {author} {\bibfnamefont {D.}~\bibnamefont {Derks}}, \ and\ \bibinfo
  {author} {\bibfnamefont {D.}~\bibnamefont {Bonn}},\ }\href@noop {} {\bibfield
   {journal} {\bibinfo  {journal} {Phil. Trans. R. Soc. Lond. A}\ }\textbf
  {\bibinfo {volume} {367}},\ \bibinfo {pages} {5139} (\bibinfo {year}
  {2009})}\BibitemShut {NoStop}%
\bibitem [{\citenamefont {Derakhshandeh}\ and\ \citenamefont
  {Vlassopoulos}(2012)}]{Derakhshandeh:2011}%
  \BibitemOpen
  \bibfield  {author} {\bibinfo {author} {\bibfnamefont {B.}~\bibnamefont
  {Derakhshandeh}}\ and\ \bibinfo {author} {\bibfnamefont {S.~G.}\ \bibnamefont
  {Vlassopoulos}, \bibfnamefont {Dimitrisand~Hatzikiriakos}},\ }\href {\doibase
  10.1007/s00397-011-0577-7} {\bibfield  {journal} {\bibinfo  {journal} {Rheol.
  Acta}\ }\textbf {\bibinfo {volume} {51}},\ \bibinfo {pages} {201} (\bibinfo
  {year} {2012})}\BibitemShut {NoStop}%
\bibitem [{\citenamefont {Divoux}\ \emph
  {et~al.}(2011{\natexlab{a}})\citenamefont {Divoux}, \citenamefont
  {Barentin},\ and\ \citenamefont {Manneville}}]{Divoux:2011b}%
  \BibitemOpen
  \bibfield  {author} {\bibinfo {author} {\bibfnamefont {T.}~\bibnamefont
  {Divoux}}, \bibinfo {author} {\bibfnamefont {C.}~\bibnamefont {Barentin}}, \
  and\ \bibinfo {author} {\bibfnamefont {S.}~\bibnamefont {Manneville}},\
  }\href@noop {} {\bibfield  {journal} {\bibinfo  {journal} {Soft Matter}\
  }\textbf {\bibinfo {volume} {7}},\ \bibinfo {pages} {8409} (\bibinfo {year}
  {2011}{\natexlab{a}})}\BibitemShut {NoStop}%
\bibitem [{\citenamefont {Vasu}\ \emph {et~al.}(2013)\citenamefont {Vasu},
  \citenamefont {Krishnaswamy}, \citenamefont {Sampath},\ and\ \citenamefont
  {Sood}}]{Vasu:2013}%
  \BibitemOpen
  \bibfield  {author} {\bibinfo {author} {\bibfnamefont {K.}~\bibnamefont
  {Vasu}}, \bibinfo {author} {\bibfnamefont {R.}~\bibnamefont {Krishnaswamy}},
  \bibinfo {author} {\bibfnamefont {S.}~\bibnamefont {Sampath}}, \ and\
  \bibinfo {author} {\bibfnamefont {A.}~\bibnamefont {Sood}},\ }\href@noop {}
  {\bibfield  {journal} {\bibinfo  {journal} {Soft Matter}\ }\textbf {\bibinfo
  {volume} {9}},\ \bibinfo {pages} {5874} (\bibinfo {year} {2013})}\BibitemShut
  {NoStop}%
\bibitem [{\citenamefont {Poumaere}\ \emph {et~al.}(2014)\citenamefont
  {Poumaere}, \citenamefont {Moyers-Gonz\'alez}, \citenamefont {Castelain},\
  and\ \citenamefont {Burghelea}}]{Poumaere:2014}%
  \BibitemOpen
  \bibfield  {author} {\bibinfo {author} {\bibfnamefont {A.}~\bibnamefont
  {Poumaere}}, \bibinfo {author} {\bibfnamefont {M.}~\bibnamefont
  {Moyers-Gonz\'alez}}, \bibinfo {author} {\bibfnamefont {C.}~\bibnamefont
  {Castelain}}, \ and\ \bibinfo {author} {\bibfnamefont {T.}~\bibnamefont
  {Burghelea}},\ }\href@noop {} {\bibfield  {journal} {\bibinfo  {journal} {J.
  Non-Newtonian Fluid Mech.}\ }\textbf {\bibinfo {volume} {205}},\ \bibinfo
  {pages} {28} (\bibinfo {year} {2014})}\BibitemShut {NoStop}%
\bibitem [{\citenamefont {Fourmentin}\ \emph {et~al.}(2015)\citenamefont
  {Fourmentin}, \citenamefont {Ovarlez}, \citenamefont {Faure}, \citenamefont
  {Peter}, \citenamefont {Lesueur}, \citenamefont {Daviller},\ and\
  \citenamefont {Coussot}}]{Fourmentin:2015}%
  \BibitemOpen
  \bibfield  {author} {\bibinfo {author} {\bibfnamefont {M.}~\bibnamefont
  {Fourmentin}}, \bibinfo {author} {\bibfnamefont {G.}~\bibnamefont {Ovarlez}},
  \bibinfo {author} {\bibfnamefont {P.}~\bibnamefont {Faure}}, \bibinfo
  {author} {\bibfnamefont {U.}~\bibnamefont {Peter}}, \bibinfo {author}
  {\bibfnamefont {D.}~\bibnamefont {Lesueur}}, \bibinfo {author} {\bibfnamefont
  {D.}~\bibnamefont {Daviller}}, \ and\ \bibinfo {author} {\bibfnamefont
  {P.}~\bibnamefont {Coussot}},\ }\href@noop {} {\bibfield  {journal} {\bibinfo
   {journal} {Rheol. Acta}\ }\textbf {\bibinfo {volume} {54}},\ \bibinfo
  {pages} {647} (\bibinfo {year} {2015})}\BibitemShut {NoStop}%
\bibitem [{\citenamefont {Mendes}\ \emph {et~al.}(2015)\citenamefont {Mendes},
  \citenamefont {Vinay}, \citenamefont {Ovarlez},\ and\ \citenamefont
  {Coussot}}]{Mendes:2015}%
  \BibitemOpen
  \bibfield  {author} {\bibinfo {author} {\bibfnamefont {R.}~\bibnamefont
  {Mendes}}, \bibinfo {author} {\bibfnamefont {G.}~\bibnamefont {Vinay}},
  \bibinfo {author} {\bibfnamefont {G.}~\bibnamefont {Ovarlez}}, \ and\
  \bibinfo {author} {\bibfnamefont {P.}~\bibnamefont {Coussot}},\ }\href@noop
  {} {\bibfield  {journal} {\bibinfo  {journal} {J. Non-Newtonian Fluid Mech.}\
  }\textbf {\bibinfo {volume} {220}},\ \bibinfo {pages} {77} (\bibinfo {year}
  {2015})}\BibitemShut {NoStop}%
\bibitem [{\citenamefont {Perret}\ \emph {et~al.}(1996)\citenamefont {Perret},
  \citenamefont {Locat},\ and\ \citenamefont {Martignoni}}]{Perret:1996}%
  \BibitemOpen
  \bibfield  {author} {\bibinfo {author} {\bibfnamefont {D.}~\bibnamefont
  {Perret}}, \bibinfo {author} {\bibfnamefont {J.}~\bibnamefont {Locat}}, \
  and\ \bibinfo {author} {\bibfnamefont {P.}~\bibnamefont {Martignoni}},\
  }\href@noop {} {\bibfield  {journal} {\bibinfo  {journal} {Eng. Geol.}\
  }\textbf {\bibinfo {volume} {43}},\ \bibinfo {pages} {31} (\bibinfo {year}
  {1996})}\BibitemShut {NoStop}%
\bibitem [{\citenamefont {Prestidge}\ \emph {et~al.}(1999)\citenamefont
  {Prestidge}, \citenamefont {Ametov},\ and\ \citenamefont
  {Addai-Mensah}}]{Prestidge:1999}%
  \BibitemOpen
  \bibfield  {author} {\bibinfo {author} {\bibfnamefont {C.}~\bibnamefont
  {Prestidge}}, \bibinfo {author} {\bibfnamefont {I.}~\bibnamefont {Ametov}}, \
  and\ \bibinfo {author} {\bibfnamefont {J.}~\bibnamefont {Addai-Mensah}},\
  }\href@noop {} {\bibfield  {journal} {\bibinfo  {journal} {Colloids Surf.,
  A}\ }\textbf {\bibinfo {volume} {157}},\ \bibinfo {pages} {137} (\bibinfo
  {year} {1999})}\BibitemShut {NoStop}%
\bibitem [{\citenamefont {Labanda}\ and\ \citenamefont
  {Llorens}(2005)}]{Labanda:2005}%
  \BibitemOpen
  \bibfield  {author} {\bibinfo {author} {\bibfnamefont {J.}~\bibnamefont
  {Labanda}}\ and\ \bibinfo {author} {\bibfnamefont {J.}~\bibnamefont
  {Llorens}},\ }\href@noop {} {\bibfield  {journal} {\bibinfo  {journal} {J.
  Colloid Interface Sci.}\ }\textbf {\bibinfo {volume} {289}},\ \bibinfo
  {pages} {86} (\bibinfo {year} {2005})}\BibitemShut {NoStop}%
\bibitem [{\citenamefont {McMillen}(1932)}]{McMillen:1932}%
  \BibitemOpen
  \bibfield  {author} {\bibinfo {author} {\bibfnamefont {E.}~\bibnamefont
  {McMillen}},\ }\href@noop {} {\bibfield  {journal} {\bibinfo  {journal} {J.
  Rheol.}\ }\textbf {\bibinfo {volume} {3}},\ \bibinfo {pages} {179} (\bibinfo
  {year} {1932})}\BibitemShut {NoStop}%
\bibitem [{\citenamefont {Green}\ and\ \citenamefont
  {Weltmann}(1948)}]{Green:1948}%
  \BibitemOpen
  \bibfield  {author} {\bibinfo {author} {\bibfnamefont {H.}~\bibnamefont
  {Green}}\ and\ \bibinfo {author} {\bibfnamefont {R.}~\bibnamefont
  {Weltmann}},\ }\href@noop {} {\bibfield  {journal} {\bibinfo  {journal} {Ind.
  End. Chem. Anal. Ed.}\ }\textbf {\bibinfo {volume} {15}},\ \bibinfo {pages}
  {201} (\bibinfo {year} {1948})}\BibitemShut {NoStop}%
\bibitem [{\citenamefont {Green}(1942)}]{Green:1942}%
  \BibitemOpen
  \bibfield  {author} {\bibinfo {author} {\bibfnamefont {H.}~\bibnamefont
  {Green}},\ }\href@noop {} {\bibfield  {journal} {\bibinfo  {journal} {Ind.
  End. Chem. Anal. Ed.}\ }\textbf {\bibinfo {volume} {14}},\ \bibinfo {pages}
  {576} (\bibinfo {year} {1942})}\BibitemShut {NoStop}%
\bibitem [{\citenamefont {Nakaishi}\ and\ \citenamefont
  {Yasutomi}(1994)}]{Nakaishi:1994}%
  \BibitemOpen
  \bibfield  {author} {\bibinfo {author} {\bibfnamefont {K.}~\bibnamefont
  {Nakaishi}}\ and\ \bibinfo {author} {\bibfnamefont {R.}~\bibnamefont
  {Yasutomi}},\ }\href@noop {} {\bibfield  {journal} {\bibinfo  {journal}
  {Appl. Clay Sci.}\ }\textbf {\bibinfo {volume} {9}},\ \bibinfo {pages} {71}
  (\bibinfo {year} {1994})}\BibitemShut {NoStop}%
\bibitem [{\citenamefont {Nakaishi}(1997)}]{Nakaishi:1997}%
  \BibitemOpen
  \bibfield  {author} {\bibinfo {author} {\bibfnamefont {K.}~\bibnamefont
  {Nakaishi}},\ }\href@noop {} {\bibfield  {journal} {\bibinfo  {journal}
  {Appl. Clay Sci.}\ }\textbf {\bibinfo {volume} {12}},\ \bibinfo {pages} {377}
  (\bibinfo {year} {1997})}\BibitemShut {NoStop}%
\bibitem [{\citenamefont {Mewis}(1979)}]{Mewis:1979}%
  \BibitemOpen
  \bibfield  {author} {\bibinfo {author} {\bibfnamefont {J.}~\bibnamefont
  {Mewis}},\ }\href@noop {} {\bibfield  {journal} {\bibinfo  {journal} {J.
  Non-Newtonian Fluid Mech.}\ }\textbf {\bibinfo {volume} {6}},\ \bibinfo
  {pages} {1} (\bibinfo {year} {1979})}\BibitemShut {NoStop}%
\bibitem [{\citenamefont {T\'arrega}\ \emph {et~al.}(2004)\citenamefont
  {T\'arrega}, \citenamefont {Dur\'an},\ and\ \citenamefont
  {Costell}}]{Tarrega:2004}%
  \BibitemOpen
  \bibfield  {author} {\bibinfo {author} {\bibfnamefont {A.}~\bibnamefont
  {T\'arrega}}, \bibinfo {author} {\bibfnamefont {L.}~\bibnamefont {Dur\'an}},
  \ and\ \bibinfo {author} {\bibfnamefont {E.}~\bibnamefont {Costell}},\ }\href
  {\doibase http://dx.doi.org/10.1016/j.idairyj.2003.12.004} {\bibfield
  {journal} {\bibinfo  {journal} {Int. Dairy J.}\ }\textbf {\bibinfo {volume}
  {14}},\ \bibinfo {pages} {345 } (\bibinfo {year} {2004})}\BibitemShut
  {NoStop}%
\bibitem [{\citenamefont {Dolz}\ \emph {et~al.}(2007)\citenamefont {Dolz},
  \citenamefont {Hern\'andez}, \citenamefont {Delegido}, \citenamefont
  {Alfaro},\ and\ \citenamefont {Mu$\tilde{n}$oz}}]{Dolz:2007}%
  \BibitemOpen
  \bibfield  {author} {\bibinfo {author} {\bibfnamefont {M.}~\bibnamefont
  {Dolz}}, \bibinfo {author} {\bibfnamefont {M.}~\bibnamefont {Hern\'andez}},
  \bibinfo {author} {\bibfnamefont {J.}~\bibnamefont {Delegido}}, \bibinfo
  {author} {\bibfnamefont {M.}~\bibnamefont {Alfaro}}, \ and\ \bibinfo {author}
  {\bibfnamefont {J.}~\bibnamefont {Mu$\tilde{n}$oz}},\ }\href {\doibase
  http://dx.doi.org/10.1016/j.jfoodeng.2006.10.020} {\bibfield  {journal}
  {\bibinfo  {journal} {J. Food Eng.}\ }\textbf {\bibinfo {volume} {81}},\
  \bibinfo {pages} {179 } (\bibinfo {year} {2007})}\BibitemShut {NoStop}%
\bibitem [{\citenamefont {Sun}\ and\ \citenamefont {Zhang}(2015)}]{Sun:2015}%
  \BibitemOpen
  \bibfield  {author} {\bibinfo {author} {\bibfnamefont {G.}~\bibnamefont
  {Sun}}\ and\ \bibinfo {author} {\bibfnamefont {J.}~\bibnamefont {Zhang}},\
  }\href@noop {} {\bibfield  {journal} {\bibinfo  {journal} {Rheol. Acta}\
  }\textbf {\bibinfo {volume} {54}},\ \bibinfo {pages} {817} (\bibinfo {year}
  {2015})}\BibitemShut {NoStop}%
\bibitem [{\citenamefont {Mewis}\ and\ \citenamefont
  {Wagner}(2009)}]{Mewis:2009}%
  \BibitemOpen
  \bibfield  {author} {\bibinfo {author} {\bibfnamefont {J.}~\bibnamefont
  {Mewis}}\ and\ \bibinfo {author} {\bibfnamefont {N.~J.}\ \bibnamefont
  {Wagner}},\ }\href@noop {} {\bibfield  {journal} {\bibinfo  {journal} {Adv.
  Colloid Interface Sci.}\ }\textbf {\bibinfo {volume} {147--148}},\ \bibinfo
  {pages} {214} (\bibinfo {year} {2009})}\BibitemShut {NoStop}%
\bibitem [{\citenamefont {Divoux}\ \emph {et~al.}(2013)\citenamefont {Divoux},
  \citenamefont {Grenard},\ and\ \citenamefont {Manneville}}]{Divoux:2013}%
  \BibitemOpen
  \bibfield  {author} {\bibinfo {author} {\bibfnamefont {T.}~\bibnamefont
  {Divoux}}, \bibinfo {author} {\bibfnamefont {V.}~\bibnamefont {Grenard}}, \
  and\ \bibinfo {author} {\bibfnamefont {S.}~\bibnamefont {Manneville}},\
  }\href@noop {} {\bibfield  {journal} {\bibinfo  {journal} {Phys. Rev. Lett.}\
  }\textbf {\bibinfo {volume} {110}},\ \bibinfo {pages} {018304} (\bibinfo
  {year} {2013})}\BibitemShut {NoStop}%
\bibitem [{\citenamefont {M{\o}ller}\ \emph {et~al.}(2008)\citenamefont
  {M{\o}ller}, \citenamefont {Rodts}, \citenamefont {Michels},\ and\
  \citenamefont {Bonn}}]{Moller:2008}%
  \BibitemOpen
  \bibfield  {author} {\bibinfo {author} {\bibfnamefont {P.~C.~F.}\
  \bibnamefont {M{\o}ller}}, \bibinfo {author} {\bibfnamefont {S.}~\bibnamefont
  {Rodts}}, \bibinfo {author} {\bibfnamefont {M.~A.~J.}\ \bibnamefont
  {Michels}}, \ and\ \bibinfo {author} {\bibfnamefont {D.}~\bibnamefont
  {Bonn}},\ }\href@noop {} {\bibfield  {journal} {\bibinfo  {journal} {Phys.
  Rev. E}\ }\textbf {\bibinfo {volume} {77}},\ \bibinfo {pages} {041507}
  (\bibinfo {year} {2008})}\BibitemShut {NoStop}%
\bibitem [{\citenamefont {Fall}\ \emph {et~al.}(2010)\citenamefont {Fall},
  \citenamefont {Paredes},\ and\ \citenamefont {Bonn}}]{Fall:2010b}%
  \BibitemOpen
  \bibfield  {author} {\bibinfo {author} {\bibfnamefont {A.}~\bibnamefont
  {Fall}}, \bibinfo {author} {\bibfnamefont {J.}~\bibnamefont {Paredes}}, \
  and\ \bibinfo {author} {\bibfnamefont {D.}~\bibnamefont {Bonn}},\ }\href@noop
  {} {\bibfield  {journal} {\bibinfo  {journal} {Phys. Rev. Lett.}\ }\textbf
  {\bibinfo {volume} {105}},\ \bibinfo {pages} {225502} (\bibinfo {year}
  {2010})}\BibitemShut {NoStop}%
\bibitem [{\citenamefont {Coussot}\ \emph {et~al.}(2009)\citenamefont
  {Coussot}, \citenamefont {Tocquer}, \citenamefont {Lanos},\ and\
  \citenamefont {Ovarlez}}]{Coussot:2009}%
  \BibitemOpen
  \bibfield  {author} {\bibinfo {author} {\bibfnamefont {P.}~\bibnamefont
  {Coussot}}, \bibinfo {author} {\bibfnamefont {L.}~\bibnamefont {Tocquer}},
  \bibinfo {author} {\bibfnamefont {C.}~\bibnamefont {Lanos}}, \ and\ \bibinfo
  {author} {\bibfnamefont {G.}~\bibnamefont {Ovarlez}},\ }\href@noop {}
  {\bibfield  {journal} {\bibinfo  {journal} {J. Non-Newtonian Fluid Mech.}\
  }\textbf {\bibinfo {volume} {158}},\ \bibinfo {pages} {85} (\bibinfo {year}
  {2009})}\BibitemShut {NoStop}%
\bibitem [{\citenamefont {Ovarlez}\ \emph {et~al.}(2010)\citenamefont
  {Ovarlez}, \citenamefont {Krishan},\ and\ \citenamefont
  {Cohen-Addad}}]{Ovarlez:2010}%
  \BibitemOpen
  \bibfield  {author} {\bibinfo {author} {\bibfnamefont {G.}~\bibnamefont
  {Ovarlez}}, \bibinfo {author} {\bibfnamefont {K.}~\bibnamefont {Krishan}}, \
  and\ \bibinfo {author} {\bibfnamefont {S.}~\bibnamefont {Cohen-Addad}},\
  }\href@noop {} {\bibfield  {journal} {\bibinfo  {journal} {Europhys. Lett.}\
  }\textbf {\bibinfo {volume} {91}},\ \bibinfo {pages} {68005} (\bibinfo {year}
  {2010})}\BibitemShut {NoStop}%
\bibitem [{\citenamefont {Divoux}\ \emph {et~al.}(2010)\citenamefont {Divoux},
  \citenamefont {Tamarii}, \citenamefont {Barentin},\ and\ \citenamefont
  {Manneville}}]{Divoux:2010}%
  \BibitemOpen
  \bibfield  {author} {\bibinfo {author} {\bibfnamefont {T.}~\bibnamefont
  {Divoux}}, \bibinfo {author} {\bibfnamefont {D.}~\bibnamefont {Tamarii}},
  \bibinfo {author} {\bibfnamefont {C.}~\bibnamefont {Barentin}}, \ and\
  \bibinfo {author} {\bibfnamefont {S.}~\bibnamefont {Manneville}},\
  }\href@noop {} {\bibfield  {journal} {\bibinfo  {journal} {Phys. Rev. Lett.}\
  }\textbf {\bibinfo {volume} {104}},\ \bibinfo {pages} {208301} (\bibinfo
  {year} {2010})}\BibitemShut {NoStop}%
\bibitem [{\citenamefont {Divoux}\ \emph
  {et~al.}(2011{\natexlab{b}})\citenamefont {Divoux}, \citenamefont
  {Barentin},\ and\ \citenamefont {Manneville}}]{Divoux:2011}%
  \BibitemOpen
  \bibfield  {author} {\bibinfo {author} {\bibfnamefont {T.}~\bibnamefont
  {Divoux}}, \bibinfo {author} {\bibfnamefont {C.}~\bibnamefont {Barentin}}, \
  and\ \bibinfo {author} {\bibfnamefont {S.}~\bibnamefont {Manneville}},\
  }\href@noop {} {\bibfield  {journal} {\bibinfo  {journal} {Soft Matter}\
  }\textbf {\bibinfo {volume} {7}},\ \bibinfo {pages} {9335} (\bibinfo {year}
  {2011}{\natexlab{b}})}\BibitemShut {NoStop}%
\bibitem [{\citenamefont {Moorcroft}\ \emph {et~al.}(2011)\citenamefont
  {Moorcroft}, \citenamefont {Cates},\ and\ \citenamefont
  {Fielding}}]{Moorcroft:2011}%
  \BibitemOpen
  \bibfield  {author} {\bibinfo {author} {\bibfnamefont {R.}~\bibnamefont
  {Moorcroft}}, \bibinfo {author} {\bibfnamefont {M.}~\bibnamefont {Cates}}, \
  and\ \bibinfo {author} {\bibfnamefont {S.}~\bibnamefont {Fielding}},\
  }\href@noop {} {\bibfield  {journal} {\bibinfo  {journal} {Phys. Rev. Lett.}\
  }\textbf {\bibinfo {volume} {106}},\ \bibinfo {pages} {055502} (\bibinfo
  {year} {2011})}\BibitemShut {NoStop}%
\bibitem [{\citenamefont {Moorcroft}\ and\ \citenamefont
  {Fielding}(2013)}]{Moorcroft:2013}%
  \BibitemOpen
  \bibfield  {author} {\bibinfo {author} {\bibfnamefont {R.}~\bibnamefont
  {Moorcroft}}\ and\ \bibinfo {author} {\bibfnamefont {S.}~\bibnamefont
  {Fielding}},\ }\href@noop {} {\bibfield  {journal} {\bibinfo  {journal}
  {Phys. Rev. Lett.}\ }\textbf {\bibinfo {volume} {110}},\ \bibinfo {pages}
  {086001} (\bibinfo {year} {2013})}\BibitemShut {NoStop}%
\bibitem [{\citenamefont {Radhakrishnan}\ and\ \citenamefont
  {Fielding}(2016)}]{Ranga_2016}%
  \BibitemOpen
  \bibfield  {author} {\bibinfo {author} {\bibfnamefont {R.}~\bibnamefont
  {Radhakrishnan}}\ and\ \bibinfo {author} {\bibfnamefont {S.~M.}\ \bibnamefont
  {Fielding}},\ }\href {\doibase 10.1103/PhysRevLett.117.188001} {\bibfield
  {journal} {\bibinfo  {journal} {Phys. Rev. Lett.}\ }\textbf {\bibinfo
  {volume} {117}},\ \bibinfo {pages} {188001} (\bibinfo {year}
  {2016})}\BibitemShut {NoStop}%
\bibitem [{\citenamefont {Cheng}\ and\ \citenamefont
  {Evans}(1965)}]{Cheng:1965}%
  \BibitemOpen
  \bibfield  {author} {\bibinfo {author} {\bibfnamefont {D.-H.}\ \bibnamefont
  {Cheng}}\ and\ \bibinfo {author} {\bibfnamefont {F.}~\bibnamefont {Evans}},\
  }\href@noop {} {\bibfield  {journal} {\bibinfo  {journal} {Brit. J. Applied .
  Phys.}\ }\textbf {\bibinfo {volume} {16}},\ \bibinfo {pages} {1599} (\bibinfo
  {year} {1965})}\BibitemShut {NoStop}%
\bibitem [{\citenamefont {Toorman}(1997)}]{Toorman:1997}%
  \BibitemOpen
  \bibfield  {author} {\bibinfo {author} {\bibfnamefont {E.}~\bibnamefont
  {Toorman}},\ }\href@noop {} {\bibfield  {journal} {\bibinfo  {journal}
  {Rheol. Acta}\ }\textbf {\bibinfo {volume} {36}},\ \bibinfo {pages} {56}
  (\bibinfo {year} {1997})}\BibitemShut {NoStop}%
\bibitem [{\citenamefont {Zhu}\ and\ \citenamefont {Smay}(2011)}]{Zhu:2011}%
  \BibitemOpen
  \bibfield  {author} {\bibinfo {author} {\bibfnamefont {C.}~\bibnamefont
  {Zhu}}\ and\ \bibinfo {author} {\bibfnamefont {J.}~\bibnamefont {Smay}},\
  }\href@noop {} {\bibfield  {journal} {\bibinfo  {journal} {J. Rheol.}\
  }\textbf {\bibinfo {volume} {55}},\ \bibinfo {pages} {655} (\bibinfo {year}
  {2011})}\BibitemShut {NoStop}%
\bibitem [{\citenamefont {{de Souza Mendes}}\ and\ \citenamefont
  {Thompson}(2012)}]{deSouzaMendes:2012}%
  \BibitemOpen
  \bibfield  {author} {\bibinfo {author} {\bibfnamefont {P.~R.}\ \bibnamefont
  {{de Souza Mendes}}}\ and\ \bibinfo {author} {\bibfnamefont {R.~L.}\
  \bibnamefont {Thompson}},\ }\href@noop {} {\bibfield  {journal} {\bibinfo
  {journal} {J. Non-Newtonian Fluid Mech.}\ }\textbf {\bibinfo {volume}
  {187--188}},\ \bibinfo {pages} {8} (\bibinfo {year} {2012})}\BibitemShut
  {NoStop}%
\bibitem [{\citenamefont {Petrellis}\ and\ \citenamefont
  {Flumerfelt}(1973)}]{Petrellis:1973}%
  \BibitemOpen
  \bibfield  {author} {\bibinfo {author} {\bibfnamefont {N.}~\bibnamefont
  {Petrellis}}\ and\ \bibinfo {author} {\bibfnamefont {R.}~\bibnamefont
  {Flumerfelt}},\ }\href@noop {} {\bibfield  {journal} {\bibinfo  {journal}
  {Can. J. Chem. Eng.}\ }\textbf {\bibinfo {volume} {51}},\ \bibinfo {pages}
  {291} (\bibinfo {year} {1973})}\BibitemShut {NoStop}%
\bibitem [{\citenamefont {Puisto}\ \emph {et~al.}(2015)\citenamefont {Puisto},
  \citenamefont {Mohtaschemi}, \citenamefont {Alava},\ and\ \citenamefont
  {Illa}}]{Puisto:2015}%
  \BibitemOpen
  \bibfield  {author} {\bibinfo {author} {\bibfnamefont {A.}~\bibnamefont
  {Puisto}}, \bibinfo {author} {\bibfnamefont {M.}~\bibnamefont {Mohtaschemi}},
  \bibinfo {author} {\bibfnamefont {M.}~\bibnamefont {Alava}}, \ and\ \bibinfo
  {author} {\bibfnamefont {X.}~\bibnamefont {Illa}},\ }\href@noop {} {\bibfield
   {journal} {\bibinfo  {journal} {Phys. Rev. E}\ }\textbf {\bibinfo {volume}
  {91}},\ \bibinfo {pages} {042314} (\bibinfo {year} {2015})}\BibitemShut
  {NoStop}%
\bibitem [{\citenamefont {Mewis}(1975)}]{Mewis1975}%
  \BibitemOpen
  \bibfield  {author} {\bibinfo {author} {\bibfnamefont {J.}~\bibnamefont
  {Mewis}},\ }\href {http://stacks.iop.org/0022-3727/8/i=12/a=005} {\bibfield
  {journal} {\bibinfo  {journal} {J. Phys. D: Appl. Phys.}\ }\textbf {\bibinfo
  {volume} {8}},\ \bibinfo {pages} {L148} (\bibinfo {year} {1975})}\BibitemShut
  {NoStop}%
\bibitem [{\citenamefont {Sainudiin}\ \emph {et~al.}(2015)\citenamefont
  {Sainudiin}, \citenamefont {Moyers-Gonzalez},\ and\ \citenamefont
  {Burghelea}}]{Sainudiin2015}%
  \BibitemOpen
  \bibfield  {author} {\bibinfo {author} {\bibfnamefont {R.}~\bibnamefont
  {Sainudiin}}, \bibinfo {author} {\bibfnamefont {M.}~\bibnamefont
  {Moyers-Gonzalez}}, \ and\ \bibinfo {author} {\bibfnamefont {T.}~\bibnamefont
  {Burghelea}},\ }\href@noop {} {\bibfield  {journal} {\bibinfo  {journal}
  {Soft Matter}\ }\textbf {\bibinfo {volume} {11}},\ \bibinfo {pages} {5531}
  (\bibinfo {year} {2015})}\BibitemShut {NoStop}%
\bibitem [{\citenamefont {Manneville}\ \emph {et~al.}(2004)\citenamefont
  {Manneville}, \citenamefont {B{\'e}cu},\ and\ \citenamefont
  {Colin}}]{Manneville:2004a}%
  \BibitemOpen
  \bibfield  {author} {\bibinfo {author} {\bibfnamefont {S.}~\bibnamefont
  {Manneville}}, \bibinfo {author} {\bibfnamefont {L.}~\bibnamefont
  {B{\'e}cu}}, \ and\ \bibinfo {author} {\bibfnamefont {A.}~\bibnamefont
  {Colin}},\ }\href@noop {} {\bibfield  {journal} {\bibinfo  {journal} {Eur.
  Phys. J. AP}\ }\textbf {\bibinfo {volume} {28}},\ \bibinfo {pages} {361}
  (\bibinfo {year} {2004})}\BibitemShut {NoStop}%
\bibitem [{\citenamefont {Gibaud}\ \emph {et~al.}(2010)\citenamefont {Gibaud},
  \citenamefont {Frelat},\ and\ \citenamefont {Manneville}}]{Gibaud:2010}%
  \BibitemOpen
  \bibfield  {author} {\bibinfo {author} {\bibfnamefont {T.}~\bibnamefont
  {Gibaud}}, \bibinfo {author} {\bibfnamefont {D.}~\bibnamefont {Frelat}}, \
  and\ \bibinfo {author} {\bibfnamefont {S.}~\bibnamefont {Manneville}},\
  }\href@noop {} {\bibfield  {journal} {\bibinfo  {journal} {Soft Matter}\
  }\textbf {\bibinfo {volume} {6}},\ \bibinfo {pages} {3482} (\bibinfo {year}
  {2010})}\BibitemShut {NoStop}%
\bibitem [{\citenamefont {Grenard}\ \emph {et~al.}(2014)\citenamefont
  {Grenard}, \citenamefont {Divoux}, \citenamefont {Taberlet},\ and\
  \citenamefont {Manneville}}]{Grenard:2014}%
  \BibitemOpen
  \bibfield  {author} {\bibinfo {author} {\bibfnamefont {V.}~\bibnamefont
  {Grenard}}, \bibinfo {author} {\bibfnamefont {T.}~\bibnamefont {Divoux}},
  \bibinfo {author} {\bibfnamefont {N.}~\bibnamefont {Taberlet}}, \ and\
  \bibinfo {author} {\bibfnamefont {S.}~\bibnamefont {Manneville}},\
  }\href@noop {} {\bibfield  {journal} {\bibinfo  {journal} {Soft Matter}\
  }\textbf {\bibinfo {volume} {10}},\ \bibinfo {pages} {1555} (\bibinfo {year}
  {2014})}\BibitemShut {NoStop}%
\bibitem [{\citenamefont {Coussot}\ \emph
  {et~al.}(2002{\natexlab{a}})\citenamefont {Coussot}, \citenamefont {Nguyen},
  \citenamefont {Huynh},\ and\ \citenamefont {Bonn}}]{Coussot:2002b}%
  \BibitemOpen
  \bibfield  {author} {\bibinfo {author} {\bibfnamefont {P.}~\bibnamefont
  {Coussot}}, \bibinfo {author} {\bibfnamefont {Q.~D.}\ \bibnamefont {Nguyen}},
  \bibinfo {author} {\bibfnamefont {H.~T.}\ \bibnamefont {Huynh}}, \ and\
  \bibinfo {author} {\bibfnamefont {D.}~\bibnamefont {Bonn}},\ }\href@noop {}
  {\bibfield  {journal} {\bibinfo  {journal} {Phys. Rev. Lett.}\ }\textbf
  {\bibinfo {volume} {88}},\ \bibinfo {pages} {175501} (\bibinfo {year}
  {2002}{\natexlab{a}})}\BibitemShut {NoStop}%
\bibitem [{\citenamefont {Ovarlez}\ \emph
  {et~al.}(2013{\natexlab{b}})\citenamefont {Ovarlez}, \citenamefont {Tocquer},
  \citenamefont {Bertrand},\ and\ \citenamefont {Coussot}}]{Ovarlez:2013}%
  \BibitemOpen
  \bibfield  {author} {\bibinfo {author} {\bibfnamefont {G.}~\bibnamefont
  {Ovarlez}}, \bibinfo {author} {\bibfnamefont {L.}~\bibnamefont {Tocquer}},
  \bibinfo {author} {\bibfnamefont {F.}~\bibnamefont {Bertrand}}, \ and\
  \bibinfo {author} {\bibfnamefont {P.}~\bibnamefont {Coussot}},\ }\href@noop
  {} {\bibfield  {journal} {\bibinfo  {journal} {Soft Matter}\ }\textbf
  {\bibinfo {volume} {9}},\ \bibinfo {pages} {5540} (\bibinfo {year}
  {2013}{\natexlab{b}})}\BibitemShut {NoStop}%
\bibitem [{\citenamefont {Armstrong}\ \emph {et~al.}(2016)\citenamefont
  {Armstrong}, \citenamefont {Beris}, \citenamefont {Rogers},\ and\
  \citenamefont {Wagner}}]{Armstrong:2016}%
  \BibitemOpen
  \bibfield  {author} {\bibinfo {author} {\bibfnamefont {M.~J.}\ \bibnamefont
  {Armstrong}}, \bibinfo {author} {\bibfnamefont {A.~N.}\ \bibnamefont
  {Beris}}, \bibinfo {author} {\bibfnamefont {S.~A.}\ \bibnamefont {Rogers}}, \
  and\ \bibinfo {author} {\bibfnamefont {N.~J.}\ \bibnamefont {Wagner}},\
  }\href@noop {} {\bibfield  {journal} {\bibinfo  {journal} {J. Rheol.}\
  }\textbf {\bibinfo {volume} {60}},\ \bibinfo {pages} {433} (\bibinfo {year}
  {2016})}\BibitemShut {NoStop}%
\bibitem [{\citenamefont {Fielding}\ \emph {et~al.}(2013)\citenamefont
  {Fielding}, \citenamefont {Moorcroft}, \citenamefont {Larson},\ and\
  \citenamefont {Cates}}]{fielding2013}%
  \BibitemOpen
  \bibfield  {author} {\bibinfo {author} {\bibfnamefont {S.~M.}\ \bibnamefont
  {Fielding}}, \bibinfo {author} {\bibfnamefont {R.~L.}\ \bibnamefont
  {Moorcroft}}, \bibinfo {author} {\bibfnamefont {R.~G.}\ \bibnamefont
  {Larson}}, \ and\ \bibinfo {author} {\bibfnamefont {M.~E.}\ \bibnamefont
  {Cates}},\ }\href@noop {} {\bibfield  {journal} {\bibinfo  {journal} {J.
  Chem. Phys.}\ }\textbf {\bibinfo {volume} {138}},\ \bibinfo {pages} {12A504}
  (\bibinfo {year} {2013})}\BibitemShut {NoStop}%
\bibitem [{\citenamefont {Lu}\ \emph {et~al.}(2000)\citenamefont {Lu},
  \citenamefont {Olmsted},\ and\ \citenamefont {Ball}}]{Lu:2000}%
  \BibitemOpen
  \bibfield  {author} {\bibinfo {author} {\bibfnamefont {C.-Y.~D.}\
  \bibnamefont {Lu}}, \bibinfo {author} {\bibfnamefont {P.~D.}\ \bibnamefont
  {Olmsted}}, \ and\ \bibinfo {author} {\bibfnamefont {R.~C.}\ \bibnamefont
  {Ball}},\ }\href@noop {} {\bibfield  {journal} {\bibinfo  {journal} {Phys.
  Rev. Lett.}\ }\textbf {\bibinfo {volume} {84}},\ \bibinfo {pages} {642}
  (\bibinfo {year} {2000})}\BibitemShut {NoStop}%
\bibitem [{\citenamefont {Olmsted}\ \emph {et~al.}(2000)\citenamefont
  {Olmsted}, \citenamefont {Radulescu},\ and\ \citenamefont
  {Lu}}]{Olmsted:2000}%
  \BibitemOpen
  \bibfield  {author} {\bibinfo {author} {\bibfnamefont {P.~D.}\ \bibnamefont
  {Olmsted}}, \bibinfo {author} {\bibfnamefont {O.}~\bibnamefont {Radulescu}},
  \ and\ \bibinfo {author} {\bibfnamefont {C.-Y.~D.}\ \bibnamefont {Lu}},\
  }\href@noop {} {\bibfield  {journal} {\bibinfo  {journal} {J. Rheol.}\
  }\textbf {\bibinfo {volume} {44}},\ \bibinfo {pages} {257} (\bibinfo {year}
  {2000})}\BibitemShut {NoStop}%
\bibitem [{\citenamefont {Press}\ \emph {et~al.}(2007)\citenamefont {Press},
  \citenamefont {Teukolsky}, \citenamefont {Vetterling},\ and\ \citenamefont
  {Flannery}}]{press1996}%
  \BibitemOpen
  \bibfield  {author} {\bibinfo {author} {\bibfnamefont {W.~H.}\ \bibnamefont
  {Press}}, \bibinfo {author} {\bibfnamefont {S.~A.}\ \bibnamefont
  {Teukolsky}}, \bibinfo {author} {\bibfnamefont {W.~T.}\ \bibnamefont
  {Vetterling}}, \ and\ \bibinfo {author} {\bibfnamefont {B.~P.}\ \bibnamefont
  {Flannery}},\ }\href@noop {} {\emph {\bibinfo {title} {Numerical Recipes 3rd
  Edition: The Art of Scientific Computing}}}\ (\bibinfo  {publisher}
  {Cambridge University Press},\ \bibinfo {address} {New York, NY, USA},\
  \bibinfo {year} {2007})\BibitemShut {NoStop}%
\bibitem [{\citenamefont {Fielding}\ \emph {et~al.}(2009)\citenamefont
  {Fielding}, \citenamefont {Cates},\ and\ \citenamefont
  {Sollich}}]{Fielding:2009}%
  \BibitemOpen
  \bibfield  {author} {\bibinfo {author} {\bibfnamefont {S.~M.}\ \bibnamefont
  {Fielding}}, \bibinfo {author} {\bibfnamefont {M.~E.}\ \bibnamefont {Cates}},
  \ and\ \bibinfo {author} {\bibfnamefont {P.}~\bibnamefont {Sollich}},\
  }\href@noop {} {\bibfield  {journal} {\bibinfo  {journal} {Soft Matter}\
  }\textbf {\bibinfo {volume} {5}},\ \bibinfo {pages} {2378} (\bibinfo {year}
  {2009})}\BibitemShut {NoStop}%
\bibitem [{\citenamefont {Voter}(2007)}]{Voter2007}%
  \BibitemOpen
  \bibfield  {author} {\bibinfo {author} {\bibfnamefont {A.~F.}\ \bibnamefont
  {Voter}},\ }\href {\doibase 10.1007/978-1-4020-5295-8_1} {\emph {\bibinfo
  {title} {Radiation Effects in Solids}}},\ Vol.\ \bibinfo {volume} {235}\
  (\bibinfo  {publisher} {Springer},\ \bibinfo {address} {Dordrecht, The
  Netherlands},\ \bibinfo {year} {2007})\ pp.\ \bibinfo {pages}
  {1--23}\BibitemShut {NoStop}%
\bibitem [{\citenamefont {Bortz}\ \emph {et~al.}(1975)\citenamefont {Bortz},
  \citenamefont {Kalos},\ and\ \citenamefont {Lebowitz}}]{bortz1975new}%
  \BibitemOpen
  \bibfield  {author} {\bibinfo {author} {\bibfnamefont {A.~B.}\ \bibnamefont
  {Bortz}}, \bibinfo {author} {\bibfnamefont {M.~H.}\ \bibnamefont {Kalos}}, \
  and\ \bibinfo {author} {\bibfnamefont {J.~L.}\ \bibnamefont {Lebowitz}},\
  }\href@noop {} {\bibfield  {journal} {\bibinfo  {journal} {J. Comput. Phys.}\
  }\textbf {\bibinfo {volume} {17}},\ \bibinfo {pages} {10} (\bibinfo {year}
  {1975})}\BibitemShut {NoStop}%
\bibitem [{\citenamefont {Gillespie}(1976)}]{gillespie1976general}%
  \BibitemOpen
  \bibfield  {author} {\bibinfo {author} {\bibfnamefont {D.~T.}\ \bibnamefont
  {Gillespie}},\ }\href@noop {} {\bibfield  {journal} {\bibinfo  {journal} {J.
  Comput. Phys.}\ }\textbf {\bibinfo {volume} {22}},\ \bibinfo {pages} {403}
  (\bibinfo {year} {1976})}\BibitemShut {NoStop}%
\bibitem [{\citenamefont {Coussot}(2014)}]{Coussot:2014}%
  \BibitemOpen
  \bibfield  {author} {\bibinfo {author} {\bibfnamefont {P.}~\bibnamefont
  {Coussot}},\ }\href@noop {} {\bibfield  {journal} {\bibinfo  {journal} {J.
  Non-Newtonian Fluid Mech.}\ }\textbf {\bibinfo {volume} {211}},\ \bibinfo
  {pages} {31} (\bibinfo {year} {2014})}\BibitemShut {NoStop}%
\bibitem [{\citenamefont {Larson}(2015)}]{Larson2015}%
  \BibitemOpen
  \bibfield  {author} {\bibinfo {author} {\bibfnamefont {R.~G.}\ \bibnamefont
  {Larson}},\ }\href {\doibase http://dx.doi.org/10.1122/1.4913584} {\bibfield
  {journal} {\bibinfo  {journal} {J. Rheol.}\ }\textbf {\bibinfo {volume}
  {59}},\ \bibinfo {pages} {595} (\bibinfo {year} {2015})}\BibitemShut
  {NoStop}%
\bibitem [{\citenamefont {Coussot}\ \emph
  {et~al.}(2002{\natexlab{b}})\citenamefont {Coussot}, \citenamefont {Raynaud},
  \citenamefont {Bertrand}, \citenamefont {Moucheront}, \citenamefont
  {Guilbaud}, \citenamefont {Huynh}, \citenamefont {Jarny},\ and\ \citenamefont
  {Lesueur}}]{Coussot:2002a}%
  \BibitemOpen
  \bibfield  {author} {\bibinfo {author} {\bibfnamefont {P.}~\bibnamefont
  {Coussot}}, \bibinfo {author} {\bibfnamefont {J.~S.}\ \bibnamefont
  {Raynaud}}, \bibinfo {author} {\bibfnamefont {F.}~\bibnamefont {Bertrand}},
  \bibinfo {author} {\bibfnamefont {P.}~\bibnamefont {Moucheront}}, \bibinfo
  {author} {\bibfnamefont {J.~P.}\ \bibnamefont {Guilbaud}}, \bibinfo {author}
  {\bibfnamefont {H.~T.}\ \bibnamefont {Huynh}}, \bibinfo {author}
  {\bibfnamefont {S.}~\bibnamefont {Jarny}}, \ and\ \bibinfo {author}
  {\bibfnamefont {D.}~\bibnamefont {Lesueur}},\ }\href@noop {} {\bibfield
  {journal} {\bibinfo  {journal} {Phys. Rev. Lett.}\ }\textbf {\bibinfo
  {volume} {88}},\ \bibinfo {pages} {218301} (\bibinfo {year}
  {2002}{\natexlab{b}})}\BibitemShut {NoStop}%
\bibitem [{\citenamefont {Ragouilliaux}\ \emph {et~al.}(2006)\citenamefont
  {Ragouilliaux}, \citenamefont {Herzhaft}, \citenamefont {Bertrand},\ and\
  \citenamefont {Coussot}}]{Ragouilliaux:2006}%
  \BibitemOpen
  \bibfield  {author} {\bibinfo {author} {\bibfnamefont {A.}~\bibnamefont
  {Ragouilliaux}}, \bibinfo {author} {\bibfnamefont {B.}~\bibnamefont
  {Herzhaft}}, \bibinfo {author} {\bibfnamefont {F.}~\bibnamefont {Bertrand}},
  \ and\ \bibinfo {author} {\bibfnamefont {P.}~\bibnamefont {Coussot}},\
  }\href@noop {} {\bibfield  {journal} {\bibinfo  {journal} {Rheol. Acta}\
  }\textbf {\bibinfo {volume} {46}},\ \bibinfo {pages} {261} (\bibinfo {year}
  {2006})}\BibitemShut {NoStop}%
\bibitem [{\citenamefont {Perge}\ \emph {et~al.}(2014)\citenamefont {Perge},
  \citenamefont {Taberlet}, \citenamefont {Gibaud},\ and\ \citenamefont
  {Manneville}}]{Perge:2014b}%
  \BibitemOpen
  \bibfield  {author} {\bibinfo {author} {\bibfnamefont {C.}~\bibnamefont
  {Perge}}, \bibinfo {author} {\bibfnamefont {N.}~\bibnamefont {Taberlet}},
  \bibinfo {author} {\bibfnamefont {T.}~\bibnamefont {Gibaud}}, \ and\ \bibinfo
  {author} {\bibfnamefont {S.}~\bibnamefont {Manneville}},\ }\href@noop {}
  {\bibfield  {journal} {\bibinfo  {journal} {J. Rheol.}\ }\textbf {\bibinfo
  {volume} {58}},\ \bibinfo {pages} {1331} (\bibinfo {year}
  {2014})}\BibitemShut {NoStop}%
\bibitem [{\citenamefont {Gibaud}\ \emph {et~al.}(2016)\citenamefont {Gibaud},
  \citenamefont {Perge}, \citenamefont {Taberlet}, \citenamefont
  {Lindstr{\"o}m},\ and\ \citenamefont {Manneville}}]{Gibaud:2016}%
  \BibitemOpen
  \bibfield  {author} {\bibinfo {author} {\bibfnamefont {T.}~\bibnamefont
  {Gibaud}}, \bibinfo {author} {\bibfnamefont {C.}~\bibnamefont {Perge}},
  \bibinfo {author} {\bibfnamefont {N.}~\bibnamefont {Taberlet}}, \bibinfo
  {author} {\bibfnamefont {S.}~\bibnamefont {Lindstr{\"o}m}}, \ and\ \bibinfo
  {author} {\bibfnamefont {S.}~\bibnamefont {Manneville}},\ }\href@noop {}
  {\bibfield  {journal} {\bibinfo  {journal} {Soft Matter}\ }\textbf {\bibinfo
  {volume} {12}},\ \bibinfo {pages} {1701} (\bibinfo {year}
  {2016})}\BibitemShut {NoStop}%
\bibitem [{\citenamefont {Montesi}\ \emph {et~al.}(2004)\citenamefont
  {Montesi}, \citenamefont {Pe${\rm \tilde{n}}$a},\ and\ \citenamefont
  {Pasquali}}]{Montesi:2004}%
  \BibitemOpen
  \bibfield  {author} {\bibinfo {author} {\bibfnamefont {A.}~\bibnamefont
  {Montesi}}, \bibinfo {author} {\bibfnamefont {A.}~\bibnamefont {Pe${\rm
  \tilde{n}}$a}}, \ and\ \bibinfo {author} {\bibfnamefont {M.}~\bibnamefont
  {Pasquali}},\ }\href@noop {} {\bibfield  {journal} {\bibinfo  {journal}
  {Phys. Rev. Lett.}\ }\textbf {\bibinfo {volume} {92}},\ \bibinfo {pages}
  {058303} (\bibinfo {year} {2004})}\BibitemShut {NoStop}%
\bibitem [{\citenamefont {Lin-Gibson}\ \emph {et~al.}(2004)\citenamefont
  {Lin-Gibson}, \citenamefont {Pathak}, \citenamefont {Grulke}, \citenamefont
  {Wang},\ and\ \citenamefont {Hobbie}}]{LinGibson:2004}%
  \BibitemOpen
  \bibfield  {author} {\bibinfo {author} {\bibfnamefont {S.}~\bibnamefont
  {Lin-Gibson}}, \bibinfo {author} {\bibfnamefont {J.~A.}\ \bibnamefont
  {Pathak}}, \bibinfo {author} {\bibfnamefont {E.~A.}\ \bibnamefont {Grulke}},
  \bibinfo {author} {\bibfnamefont {H.}~\bibnamefont {Wang}}, \ and\ \bibinfo
  {author} {\bibfnamefont {E.~K.}\ \bibnamefont {Hobbie}},\ }\href@noop {}
  {\bibfield  {journal} {\bibinfo  {journal} {Phys. Rev. Lett.}\ }\textbf
  {\bibinfo {volume} {92}},\ \bibinfo {pages} {048302} (\bibinfo {year}
  {2004})}\BibitemShut {NoStop}%
\bibitem [{\citenamefont {Osuji}\ and\ \citenamefont
  {Weitz}(2008)}]{Osuji2008}%
  \BibitemOpen
  \bibfield  {author} {\bibinfo {author} {\bibfnamefont {C.~O.}\ \bibnamefont
  {Osuji}}\ and\ \bibinfo {author} {\bibfnamefont {D.~A.}\ \bibnamefont
  {Weitz}},\ }\href@noop {} {\bibfield  {journal} {\bibinfo  {journal} {Soft
  Matter}\ }\textbf {\bibinfo {volume} {4}},\ \bibinfo {pages} {1388} (\bibinfo
  {year} {2008})}\BibitemShut {NoStop}%
\bibitem [{\citenamefont {Grenard}\ \emph {et~al.}(2011)\citenamefont
  {Grenard}, \citenamefont {Taberlet},\ and\ \citenamefont
  {Manneville}}]{Grenard:2011}%
  \BibitemOpen
  \bibfield  {author} {\bibinfo {author} {\bibfnamefont {V.}~\bibnamefont
  {Grenard}}, \bibinfo {author} {\bibfnamefont {N.}~\bibnamefont {Taberlet}}, \
  and\ \bibinfo {author} {\bibfnamefont {S.}~\bibnamefont {Manneville}},\
  }\href@noop {} {\bibfield  {journal} {\bibinfo  {journal} {Soft Matter}\
  }\textbf {\bibinfo {volume} {7}},\ \bibinfo {pages} {3920} (\bibinfo {year}
  {2011})}\BibitemShut {NoStop}%
\bibitem [{\citenamefont {Cates}\ and\ \citenamefont
  {Sollich}(2004)}]{Cates:2004}%
  \BibitemOpen
  \bibfield  {author} {\bibinfo {author} {\bibfnamefont {M.~E.}\ \bibnamefont
  {Cates}}\ and\ \bibinfo {author} {\bibfnamefont {P.}~\bibnamefont
  {Sollich}},\ }\href@noop {} {\bibfield  {journal} {\bibinfo  {journal} {J.
  Rheol.}\ }\textbf {\bibinfo {volume} {48}},\ \bibinfo {pages} {193} (\bibinfo
  {year} {2004})}\BibitemShut {NoStop}%
\end{thebibliography}%
\end{document}